\documentclass[12pt]{article}
\usepackage{amsfonts,mathrsfs}
\begin{document}
\begin{center}
{\Large\bf Canonical description of $D=10$ superstring formulated in supertwistor space}\\[0.5cm]
{\large D.V.~Uvarov\footnote{E-mail: d\_uvarov@\,hotmail.com, uvarov@\,kipt.kharkov.ua}}\\[0.4cm]
{\it National Science Center\\ Kharkov Institute of Physics and Technology,}\\ {\it 61108 Kharkov, Ukraine}\\[0.5cm]
\end{center}

\begin{abstract}
\noindent Canonical description of the $D=10$ superstring action
involving supertwistor variables generalizing Penrose-Ferber
supertwistors is developed. Primary and secondary constraints are
identified and arranged into the first- and second-class sets.
Dirac brackets are introduced and the deformation of the Poisson
bracket algebra of the first-class constraints is studied. The
role of the deformation parameter is played by $\alpha'$.
\end{abstract}
\section{Introduction}

Twistors \cite{Penrose} and supertwistors \cite{Ferber} are known
to find one of the interesting applications in describing the
models of point-like and extended relativistic objects
\cite{Shirafuji}-\cite{FedLuk}. Such (super)twistor formulations
based on the introduction of commuting spinor variables represent a 
valuable alternative to the conventional (super)space formulation
and allow to overcome the problem of handling $\kappa-$symmetry
and streamline the covariant quantization in the case of
(super)particle models. 
Thus incorporating supertwistors into the string theory could also be useful in quest of a solution of 
the long-standing problem of Green-Schwarz (GS) superstring covariant quantization. However, 
twistor description of (supersymmetric)
models of extended objects attracted much less attention until the
twistor strings \cite{Witten}, \cite{Berkovits}\footnote{See also
\cite{Siegel04}-\cite{MShet} and \cite{Popov}-\cite{AZHM} for
other twistor string models corresponding to $4d$ gauge theories and
supergravities.} have been proposed in the context of gauge
fields/string correspondence. 
Although twistor string models appear to be
interesting objects for study and stimulated recent progress in
perturbative (super-)Yang-Mills and gravitation \cite{CS} they differ from the GS
superstrings.

In recent years the progress in solving the problem of superstring covariant quantization was mainly due to Berkovits formalism \cite{PSB2000}. Its key ingerdient is the BRST operator involving 10-dimensional pure spinor field with commuting complex components that plays the role of the ghost field for the fermionic constraints of the GS superstring and can also be viewed as the half of twistor \cite{Cherkis}. The main advantage of the Berkovits approach is the possibility to find covariant expressions for the string scattering amplitudes \cite{PSB2000}, \cite{tree}, \cite{multiloop}. It is worthwhile to note that originally pure spinors appeared as the world-sheet superpartners of superspace Grassmann coordinates $\theta^\alpha$ in the heterotic string formulation \cite{Tonin}, \cite{Berkovits92} with $n=2$ local world-sheet supersymmetry\footnote{Pure spinors also turned out to be useful in the study of the geometry of superfield constraints in super-Yang-Mills and supergravity theories \cite{Nilsson}, \cite{Howe91}.}. In the formulation of Refs. \cite{Tonin}, \cite{Berkovits92} pure spinors can also be interpreted as the pair of elements from the basis in the auxiliary spinor space extending $D=10$ $N=1$ superspace. An interesting problem of finding the relation between the Berkovits model and the GS superstrings or their classically equivalent reformulations is being investigated since the year 2000 (see \cite{Berkovits08} and references therein).

Taking into account the above mentioned results that indicate important role played by spinors/twistors in the quantum theory of superparticles and superstrings we have started to study the twistor
formulation for GS superstrings in dimensions D=4,6,10 \cite{CQG06}, \cite{CQG07} aiming at getting novel insights into the covariant qunatization problem.  Since commuting spinor variables, the necessary ingredients of twistors,
are absent in the original GS superstring action we considered,
similarly to the twistor transform procedure for (super)particles,
the classically equivalent first-order superstring action \cite{BZstring},  
where such bosonic spinors form the basis in auxiliary space -- the space of Lorentz harmonics \cite{GIKOS}-\cite{harmonics}. 
Note that $D=4$ spinor Lorentz harmonics are nothing but the normalized
Newman-Penrose dyad \cite{NP}. The presence of Lorentz harmonic variables in the Lagrangian allows to realize
$\kappa-$symmetry transformations in the irreducible form. In the formulations of \cite{Tonin}, \cite{Berkovits92} pure spinors in the similar way provide   irreducible realization of the part of $\kappa-$symmetries. Detailed discussion of the relation between the superstring formulations involving Lorentz harmonics and those with local world-sheet supersymmetry can be found in \cite{Sreview}. 

The supertwistors appearing after the twistor transform of the first-order action \cite{BZstring} 
coincide for the
$D=4$ case with those introduced by Ferber
\cite{Ferber}\footnote{Recently alternative to Ferber construction of the supertwistors has been proposed \cite{Z07},
where only conformal superPoincare symmetry is manifest and odd
supertwistor components are given by the complex Lorentz vectors
related to those appearing in the particle and string models with
world-line/world-sheet supersymmetry. Since superstring
Lagrangian, due to the presence of the dimensionful tension
parameter, is not invariant under conformal transformations it
could be of interest to consider its formulation in terms of such
alternative supertwistors or their higher-dimensional
generalizations \cite{Berkovits91}, \cite{harmonics}.}, while in
higher dimensions \cite{Bars}, \cite{BdAM} they realize the
fundamental representation of the (generalized) superconformal
group, include spinor harmonics as their projectional parts and
the Grassmann-odd components of supertwistors are represented by
the Lorentz scalars that is attractive feature from the
perspective of fixing the gauge freedom related to the
$\kappa-$symmetry. In \cite{CQG06}, \cite{CQG07} there was found
the supertwistor representation for $D=4,6,10$ superstring
Lagrangian characterized by the nondegenerate kinetic term for the
supertwistor components, derived the equations of motion, and
obtained the supertwistor realization of the $\kappa-$symmetry
transformations.

Since the superstring Lagrangian after the twistor transform is nonlinear like in the
space-time formulation and the supertwistors are constrained
variables the canonical formalism appears to be the most suitable
one for further investigation. That is why here we move to the
canonical description of the $D=10$ superstring model formulated
in terms of supertwistors. In Section 2 we identify the constraints that arise
in the process of transition to the Hamiltonian formulation and
classify them on the first- and second-class ones. Up to that step
our consideration can be viewed as the twistor counterpart of the
canonical treatment of Lorentz-harmonic superstring in the
superspace formulation \cite{BZstring}. Then in Section 3 we proceed to propose
the basis for the second-class constraints for which the Dirac
matrix acquires block diagonal structure on the constraint shell,
introduce the Dirac brackets (D.B.)
 and evaluate D.B. algebra of the
first-class constraints.

\section{Total Hamiltonian and the first-class constraints}

Based of the classification of superconformal algebras in various dimensions \cite{vHvP} it was suggested in \cite{BdAM}, \cite{CQG07} to define $D=10$ supertwistor as transforming in the fundamental representation of the $OSp(32|1)$ supergroup
\begin{equation}\label{twistor10}
Z^{\mathbf\Lambda}=(\mu^{\alpha},
v_{\alpha}, \eta).
\end{equation}
Thus it is composed of the primary spinor $\mu^{\alpha}$ and
projectional $v_\alpha$ parts that are 16-component MW spinors of
opposite chiralities and the Grassmann-odd scalar $\eta$. Such
definition generalizes the basic property of Ferber supertwistors
\cite{Ferber} to realize the fundamental representation of
$SU(2,2|N)$ locally isomorphic to the $N-$extended superconformal
group in 4 dimensions. Twistor transform for the $D=10$
superstring in the formulation with irreducible realization of the
$\kappa-$symmetry leads one to consider two sets of the
supertwistors
\begin{equation}\label{twistor10str}
Z^{\mathbf\Lambda+}_{A}=(\mu^{\alpha+}_{A},
v^+_{\alpha A}, \eta^+_{A}),\quad Z^{\mathbf\Lambda-}_{\dot A}=(\mu^{\alpha-}_{\dot A},
v^-_{\alpha\dot A}, \eta^-_{\dot A}),
\end{equation}
whose projectional parts are identified with the spinor harmonic
matrix $v^{(\alpha)}_\alpha=(v^+_{\alpha A}, v{}^-_{\alpha\dot
A})\in Spin(1,9)$ decomposed into two blocks carrying $SO(1,1)$ indices $\pm$
and transforming in the spinor representations of $Spin(8)$ $A,\dot
A=1,...,8$ in accordance with that the embedding of the string world-sheet
into the $D=10$ space-time spontaneously breaks $SO(1,9)$
symmetry down to $SO(1,1)\times SO(8)$. The $D=10$
generalization of the Penrose-Ferber incidence relations\footnote{Such $D=10$ incidence relations with $X^{\alpha\beta}$ matrix involving solely space-time coordinates $x^m$ contribution have been proposed in \cite{harmonics}.}
\begin{equation}
\begin{array}{c}
\mu^{\alpha+}_{A}=(X^{\alpha\beta}-8i\theta^{\alpha}\theta^{\beta})v^+_{\beta A},\quad\eta^+_{A}=4v^+_{\alpha
A}\theta^{\alpha},\\[0.2cm]
\mu^{\alpha-}_{\dot A}=(X^{\alpha\beta}-8i\theta^{\alpha}\theta^{\beta})v^-_{\beta\dot A},\quad\eta^-_{\dot A}=4v^-_{\alpha\dot A}\theta^{\alpha}
\end{array}
\end{equation}
involves arbitrary $16\times 16$ matrix
$X^{\alpha\beta}=x^m\tilde\sigma^{\alpha\beta}_m+z^{m_1m_2m_3}\tilde\sigma_{m_1m_2m_3}^{\alpha\beta}+z^{m_1...m_5}\tilde\sigma^{\alpha\beta}_{m_1...m_5}$
that contains, except for $D=10$ Minkowski coordinates $x^m$, antisymmetric tensor coordinates $z^{m_1m_2m_3}$ and
$z^{m_1...m_5}$ associated with tensor generators of $OSp(32|1)$. Since our goal is to describe the superstring in $D=10$ Minkowski superspace the dependence on such tensor coordinates has to be removed by the constraints
\begin{equation}\label{x3}
N^{+2}_{AB}=Z^{\mathbf\Lambda+}_AG_{\mathbf{\Lambda\Sigma}}Z^{\mathbf\Sigma+}_B\approx0,\ N^{-2}_{\dot A\dot B}=Z^{\mathbf\Lambda-}_{\dot
A}G_{\mathbf{\Lambda\Sigma}}Z^{\mathbf\Sigma-}_{\dot
B}\approx0,\ N_{A\dot A}=Z^{\mathbf\Lambda+}_AG_{\mathbf{\Lambda\Sigma}}Z^{\mathbf\Sigma-}_{\dot
A}\approx0,
\end{equation}
where
\begin{equation}
G_{\mathbf{\Lambda\Sigma}}=\left(\begin{array}{ccc}
0& \delta_{\alpha}^{\beta}& 0\\[0.2cm]
-\delta_{\beta}^{\alpha}& 0& 0\\[0.2cm]
0 & 0 &  -i\\
\end{array}
\right)
\end{equation}
is the $OSp(32|1)$ invariant orthosymplectic metric,
and
\begin{equation}\label{x5}
N_{m_1...m_5}=\sigma_{m_1...m_5\alpha\beta}(\mu^{\alpha+}_Av^{\beta-}_A+\mu^{\alpha-}_{\dot
A}v^{\beta+}_{\dot A})\approx0.
\end{equation}
The latter constraints involve inverse spinor harmonic matrix
\begin{equation}
v^{\alpha}_{(\alpha)}=(v^{\alpha-}_A, v^{\alpha+}_{\dot A}):\quad
v^{\alpha}_{(\alpha)}v_{\alpha}^{(\beta)}=\delta_{(\alpha)}^{(\beta)}.
\end{equation}

The first-order action of the $D=10$ $N=1$ superstring reformulated in
terms of supertwistors (\ref{twistor10str}) was found in \cite{CQG07}. For the
transition to the canonical  formulation it is convenient instead of
the zweibein $e^{\pm2}_\mu(\xi)$ and its inverse
$e^{\mu\pm2}(\xi)$, upon which the action depends nonlinearly, to
introduce world-sheet vector densities
$\rho^{\pm2}=(\alpha')^{1/2}ee^{\mu\pm2}$, $e=det(e^{\pm2}_\mu)$ \cite{BZstring} so that
the superstring action in the twistor formulation acquires the form
\begin{equation}\label{action}
S=-{\textstyle\frac{1}{2\alpha'}}\int d^2\xi(\rho^{\mu+2}\omega^{-2}_\mu+\rho^{\mu-2}\omega^{+2}_\mu+\varepsilon_{\mu\nu}\rho^{\mu-2}\rho^{\nu+2})+S_{WZ},
\end{equation}
where $S_{WZ}$ is the WZ term in the twistor representation given by
\begin{equation}\label{WZ}
S_{WZ}={\textstyle\frac{is}{\alpha'}}\int d^2\xi({\textstyle\frac12}\omega^{+2}(d)\wedge\varphi^{-2}(d)+{\textstyle\frac12}\omega^{-2}(d)\wedge\varphi^{+2}(d)-\omega^{I}(d)\wedge\varphi^{I}(d)).
\end{equation}
In (\ref{action}), (\ref{WZ}) $\xi^\mu=(\tau,\sigma)$ are the world-sheet local coordinates, $\alpha'$ is the Regge slope parameter, and $s=\pm1$ indicates the arbitrariness in the definition of the WZ part of the action.
The action depends on the world-sheet projections of the 1-forms
\begin{equation}\label{bridge10}
\begin{array}{c}
\omega^{+2}(d)=
{\textstyle\frac18}dZ^{\mathbf\Lambda+}_AG_{\mathbf{\Lambda\Sigma}}Z^{\mathbf\Sigma+}_A,\quad
\omega^{-2}(d)={\textstyle\frac18}dZ^{\mathbf\Lambda-}_{\dot
A}G_{\mathbf{\Lambda\Sigma}}Z^{\mathbf\Sigma-}_{\dot A},\\[0.2cm]
\omega^{I}(d)={\textstyle\frac{1}{16}}\gamma^{I}_{A\dot
A}(dZ^{\mathbf\Lambda+}_AG_{\mathbf{\Lambda\Sigma}}Z^{\mathbf{\Sigma}-}_{\dot
A}+dZ^{\mathbf\Lambda-}_{\dot
A}G_{\mathbf{\Lambda\Sigma}}Z^{\mathbf\Sigma+}_A),
\end{array}
\end{equation}
and also
\begin{equation}\label{phi10}
\varphi^{+2}(d)={\textstyle\frac18}{\cal
D}\eta^+_A\eta^+_A,\quad\varphi^{-2}(d)={\textstyle\frac18}{\cal D}\eta^-_{\dot
A}\eta^-_{\dot A},\quad\varphi^{I}(d)={\textstyle\frac{1}{16}}\gamma^{I}_{A\dot
A}({\cal D}\eta^+_A\eta^-_{\dot A}+{\cal D}\eta^-_{\dot
A}\eta^+_A).
\end{equation}
$SO(1,9)$ covariant differentials of the odd supertwistor
components are defined as
\begin{equation}
\begin{array}{c}
{\cal D}\eta^+_A=d\eta^+_A+\frac14\Omega^{+2-2}(d)\eta^+_A-\frac12\Omega^{+2I}(d)\gamma^{I}_{A\dot A}\eta^-_{\dot A}-\frac14\Omega^{IJ}(d)\gamma^{IJ}_{AB}\eta^{+}_B,\\[0.2cm]
{\cal D}\eta^-_{\dot A}=d\eta^-_{\dot
A}-\frac14\Omega^{+2-2}(d)\eta^-_{\dot
A}-\frac12\Omega^{-2I}(d)\tilde\gamma^{I}_{\dot
AA}\eta^+_{A}-\frac14\Omega^{IJ}(d)\tilde\gamma^{IJ}_{\dot
A\dot B}\eta^{-}_{\dot B},
\end{array}
\end{equation}
where $\gamma^I_{A\dot A}$ are $8d$ chiral $\gamma$-matrices,
satisfying the condition $\gamma^I_{A\dot B}\tilde\gamma^J_{\dot
BB}+(I\leftrightarrow J)=2\delta^{IJ}\delta_{AB}$,
$\gamma^{IJ}_{AB}=\frac12(\gamma^I_{A\dot A}\tilde\gamma^{J}_{\dot
AB}-\gamma^J_{A\dot A}\tilde\gamma^{I}_{\dot AB})$ and 
$\tilde\gamma^{IJ}_{\dot A\dot B}=\frac12(\tilde\gamma^I_{\dot
AA}\gamma^{J}_{A\dot B}-\tilde\gamma^J_{\dot AA}\gamma^{I}_{A\dot
B})$ are the $Spin(8)$ generators in the $c$ and $s$
representations, and contain $SO(1,1)\times SO(8)$ split
components of the trivial $SO(1,9)$ connection constructed
out of the spinor harmonics $v^{(\alpha)}_\alpha=(v^{+}_{\alpha A}, v^-_{\alpha\dot
A})$ and their inverse $v^\alpha_{(\alpha)}=(v^{\alpha-}_A, v^{\alpha+}_{\dot A})$
\begin{equation}\label{Cartanforms}
\begin{array}{c}
\Omega^{+2-2}(d)=\frac14(dv^-_{\alpha\dot A}v^{\alpha+}_{\dot A}-dv^+_{\alpha A}v^{\alpha-}_A),\\[0.2cm]
\Omega^{+2I}(d)=\frac14dv^+_{\alpha A}\gamma^{I}_{A\dot A}v^{\alpha+}_{\dot A},\ \Omega^{-2I}(d)=\frac14dv^-_{\alpha\dot A}\tilde\gamma^{I}_{\dot AA}v^{\alpha-}_A,\\[0.2cm]
\Omega^{IJ}(d)=\frac18(dv^+_{\alpha
A}\gamma^{IJ}_{AB}v^{\alpha-}_{B}+dv^-_{\alpha\dot
A}\tilde\gamma^{IJ}_{\dot A\dot
B}v^{\alpha+}_{\dot B}).
\end{array}
\end{equation}

Moving to the canonical formulation we introduce the momenta
densities
\begin{equation}
P_{\mathfrak M}(\tau,\sigma)=\frac{\vec\delta S}{\delta\dot
Q^{\mathfrak M}(\tau,\sigma)}=\left\{p_{(\mu)\alpha A}^-,
p_{(\mu)\alpha\dot A}^+, p^{\alpha-}_{(v)A}, p^{\alpha+}_{(v)\dot
A}, \pi^-_A, \pi^+_{\dot A}, p^{+}_{(v)\alpha A},
p^{-}_{(v)\alpha\dot A}, P^{\pm2}_\mu\right\}
\end{equation}
conjugate to the string coordinates
\begin{equation}
Q^{\mathfrak M}(\tau,\sigma)=\left\{\mu^{\alpha+}_{A},
\mu^{\alpha-}_{\dot A}, v^+_{\alpha A}, v^-_{\alpha\dot A},
\eta^+_A, \eta^-_{\dot A}, v^{\alpha-}_A, v^{\alpha+}_{\dot A},
\rho^{\mu\pm2}\right\}
\end{equation}
 on the Poisson brackets (P.B.)
\begin{equation}
\{P_{\mathfrak M}(\sigma),Q^{\mathfrak N}(\sigma')\}=\delta_{\mathfrak M}^{\mathfrak N}\delta(\sigma-\sigma').
\end{equation}
From the definition of momenta densities conjugate to primary
spinor parts of supertwistors there follow the constraints
\begin{equation}\label{mumom}
\begin{array}{c}
\Phi^-_{\alpha A}(\sigma)=p_{(\mu)\alpha A}^-+\frac{1}{16\alpha'}(\rho^{\tau-2}-is\varphi^{-2}_\sigma)v^+_{\alpha A}+\frac{is}{16\alpha'}\varphi^I_\sigma\gamma^I_{A\dot A}v^-_{\alpha\dot A}\approx0,\\[0.2cm]
\Phi^+_{\alpha\dot A}(\sigma)=p_{(\mu)\alpha\dot A}^++\frac{1}{16\alpha'}(\rho^{\tau+2}-is\varphi^{+2}_\sigma)v^-_{\alpha\dot A}+\frac{is}{16\alpha'}\varphi^I_\sigma\tilde\gamma^I_{\dot AA}v^+_{\alpha A}\approx0
\end{array}
\end{equation}
and analogously from the definition of momenta densities conjugate
to anticommuting supertwistor components there stem the fermionic
constraints that can be presented as
\begin{equation}\label{etamom}
\begin{array}{c}
D^-_A(\sigma)=\pi^-_A+\frac{1}{16\alpha'}(is\omega^{-2}_\sigma-s\varphi^{-2}_\sigma-i\rho^{\tau-2})\eta^+_{A}
+\frac{s}{16\alpha'}(\varphi^I_\sigma-i\omega^I_\sigma)\gamma^I_{A\dot A}\eta^-_{\dot A}\\[0.2cm]
+\frac{i}{2}\eta^+_B(v^{\alpha-}_Bp^-_{(\mu)\alpha A}-v^{\alpha-}_Ap^-_{(\mu)\alpha B})+\frac{i}{2}\eta^-_{\dot B}(v^{\alpha+}_{\dot B}p^-_{(\mu)\alpha A}-v^{\alpha-}_Ap^+_{(\mu)\alpha\dot B})\approx0,\\[0.2cm]
D^+_{\dot A}(\sigma)=\pi^+_{\dot
A}+\frac{1}{16\alpha'}(is\omega^{+2}_\sigma-s\varphi^{+2}_\sigma-i\rho^{\tau+2})\eta^-_{\dot
A}
+\frac{s}{16\alpha'}(\varphi^I_\sigma-i\omega^I_\sigma)\tilde\gamma^I_{\dot
AA}\eta^+_{A}\\[0.2cm]
+\frac{i}{2}\eta^+_B(v^{\alpha-}_Bp^+_{(\mu)\alpha\dot
A}-v^{\alpha+}_{\dot A}p^-_{(\mu)\alpha
B})+\frac{i}{2}\eta^-_{\dot B}(v^{\alpha+}_{\dot
B}p^+_{(\mu)\alpha\dot A}-v^{\alpha+}_{\dot A}p^+_{(\mu)\alpha\dot
B})\approx0.
\end{array}
\end{equation}
In the harmonic sector one finds the primary constraints arising from
the definition of momenta conjugate to spinor harmonics $v^+_{\alpha A}$ and
$v^-_{\alpha\dot A}$
\begin{equation}\label{hmom-}
\begin{array}{c}
T^{\alpha-}_A(\sigma)=p^{\alpha-}_{(v)A}+\frac{1}{16\alpha'}(is\varphi^{-2}_\sigma-\rho^{\tau-2})\mu^{\alpha+}_A
-\frac{is}{16\alpha'}\varphi^I_\sigma\gamma^I_{A\dot
A}\mu^{\alpha-}_{\dot
A}+\frac{is}{128\alpha'}\omega^I_\sigma(\eta^+\gamma^I\eta^-)v^{\alpha-}_A\\[0.2cm]
+\frac{is}{256\alpha'}\left[\frac{1}{2}\omega^{-2}_\sigma(\eta^+\gamma^{IJ}\eta^+)+\frac{1}{2}\omega^{+2}_\sigma(\eta^-\tilde\gamma^{IJ}\eta^-)
-\omega^K_\sigma(\eta^+\gamma^{IJK}\eta^-)\right]\gamma^{IJ}_{AB}v^{\alpha-}_B\\[0.2cm]
+\frac{is}{128\alpha'}[\omega^{-2}_\sigma(\eta^+\gamma^I\eta^-)+\omega^J_\sigma(\eta^-\tilde\gamma^{IJ}\eta^-)]\gamma^I_{A\dot
A}v^{\alpha+}_{\dot A}\approx0,
\end{array}
\end{equation}
\begin{equation}\label{hmom+}
\begin{array}{c}
T^{\alpha+}_{\dot A}(\sigma)=p^{\alpha+}_{(v)\dot A}+\frac{1}{16\alpha'}(is\varphi^{+2}_\sigma-\rho^{\tau+2})\mu^{\alpha-}_{\dot A}
-\frac{is}{16\alpha'}\varphi^I_\sigma\tilde\gamma^I_{\dot
AA}\mu^{\alpha+}_{A}-\frac{is}{128\alpha'}\omega^I_\sigma(\eta^+\gamma^I\eta^-)v^{\alpha+}_{\dot A}\\[0.2cm]
+\frac{is}{256\alpha'}\left[\frac{1}{2}\omega^{-2}_\sigma(\eta^+\gamma^{IJ}\eta^+)+\frac{1}{2}\omega^{+2}_\sigma(\eta^-\tilde\gamma^{IJ}\eta^-)
-\omega^K_\sigma(\eta^+\gamma^{IJK}\eta^-)\right]\tilde\gamma^{IJ}_{\dot A\dot B}v^{\alpha+}_{\dot B}\\[0.2cm]
-\frac{is}{128\alpha'}[\omega^{+2}_\sigma(\eta^+\gamma^I\eta^-)-\omega^J_\sigma(\eta^+\gamma^{IJ}\eta^+)]\tilde\gamma^I_{\dot
AA}v^{\alpha-}_{A}\approx0
\end{array}
\end{equation}
and their inverse $v^{\alpha-}_A$, $v^{\alpha+}_{\dot A}$
\begin{equation}\label{hmominverse}
p^{+}_{(v)\alpha A}(\sigma)\approx0,\quad p^{-}_{(v)\alpha\dot A}(\sigma)\approx0.
\end{equation}
The momenta densities for $\rho^{\mu\pm2}$ also enter the set of primary
constraints
\begin{equation}\label{rhotau}
P^{\pm2}_\tau(\sigma)\approx0,
\end{equation}
\begin{equation}\label{rhosigma}
P^{\pm2}_\sigma(\sigma)\approx0.
\end{equation}

Besides that $16\times16$ spinor harmonic matrix $v^{(\alpha)}_\alpha$ is constrained by 211 relations \cite{harmonics}, \cite{BZstring}
\begin{equation}\label{harm1}
\begin{array}{c}
n_m^{(k)}v^{(\alpha)}_\alpha\tilde\sigma^{mm_1...m_4\alpha\beta}v^{(\beta)}_\beta\sigma_{(k)(\alpha)(\beta)}\approx0,\\[0.2cm]
n^{+2}_mn^{m-2}-2={\textstyle\frac{1}{64}}(v^+_{\alpha
A}\tilde\sigma^{\alpha\beta}_mv^+_{\beta A})(v^-_{\gamma\dot
A}\tilde\sigma^{m\gamma\delta}v^-_{\delta\dot A})-2\approx0
\end{array}
\end{equation}
reducing its contents to 45 independent components equal to the dimension of the Spin(1,9) group.
Defining relations for the inverse harmonics, when considered as independent degrees of freedom,
\begin{equation}\label{harm2}
v^{(\alpha)}_\alpha
v^\alpha_{(\beta)}-\delta^{(\alpha)}_{(\beta)}\approx0
\end{equation}
also should be treated as constraints, as well as, the twistor constraints (\ref{x3}), (\ref{x5}).

It was shown in \cite{BZstring} that the canonical analysis
simplifies essentially if one excludes from the set of constraints
harmonicity conditions (\ref{harm1}), (\ref{harm2}) and
appropriate projections of the harmonic momenta
(\ref{hmom-})-(\ref{hmominverse}), forming on P.B. conjugate pairs
of the second-class constraints, by introducing corresponding D.B.
A suggestive feature of the D.B. is that they coincide with the
P.B. for the subspace of the phase-space defined by the primary
constraints (\ref{mumom}), (\ref{etamom}) and (\ref{rhotau}),
(\ref{rhosigma}) and the projections of harmonic momenta
\begin{equation}\label{m+2-2}
\begin{array}{c}
M^{+2-2}(\sigma)=v^+_{\alpha A}p^{\alpha-}_{(v)A}-v^-_{\alpha\dot
A}p^{\alpha+}_{(v)\dot A}-v^{\alpha-}_Ap^+_{(v)\alpha A}
+v^{\alpha+}_{\dot A}p^-_{(v)\alpha\dot
A}+\mu^{\alpha+}_Ap^-_{(\mu)\alpha A}-\mu^{\alpha-}_{\dot
A}p^+_{(\mu)\alpha\dot A}
\\[0.2cm]
+\eta^+_A\pi^-_A-\eta^-_{\dot A}\pi^+_{\dot A}\approx0,
\end{array}
\end{equation}
\begin{equation}\label{m+2i}
M^{+2I}(\sigma)=-v^+_{\alpha A}\gamma^I_{A\dot
A}p^{\alpha+}_{(v)\dot A}+v^{\alpha+}_{\dot A}\tilde\gamma^I_{\dot
AA}p^+_{(v)\alpha
A}-\mu^{\alpha+}_A\gamma^I_{A\dot A}p^+_{(\mu)\alpha\dot A}-\eta^+_A\gamma^I_{A\dot A}\pi^+_{\dot A}\approx0,
\end{equation}
\begin{equation}\label{m-2i}
M^{-2I}(\sigma)=-v^-_{\alpha\dot A}\tilde\gamma^I_{\dot
AA}p^{\alpha-}_{(v)A}+v^{\alpha-}_{A}\gamma^I_{A\dot
A}p^-_{(v)\alpha\dot A}
-\mu^{\alpha-}_{\dot A}\tilde\gamma^I_{\dot AA}p^-_{(\mu)\alpha A}-\eta^-_{\dot A}\tilde\gamma^I_{\dot AA}\pi^-_A\approx0,
\end{equation}
\begin{equation}\label{mij}
\begin{array}{c}
M^{IJ}(\sigma)=-\frac12(v^+_{\alpha
A}\gamma^{IJ}_{AB}p^{\alpha-}_{(v)B}+v^-_{\alpha\dot
A}\tilde\gamma^{IJ}_{\dot A\dot B}p^{\alpha+}_{(v)\dot B}
+v^{\alpha-}_A\gamma^{IJ}_{AB}p^+_{(v)\alpha
B}+v^{\alpha+}_{\dot A}\tilde\gamma^{IJ}_{\dot A\dot
B}p^-_{(v)\alpha\dot
B}\\[0.2cm]
+\mu^{\alpha+}_A\gamma^{IJ}_{AB}p^-_{(\mu)\alpha
B}+\mu^{\alpha-}_{\dot A}\tilde\gamma^{IJ}_{\dot A\dot
B}p^+_{(\mu)\alpha \dot
B}+\eta^+_A\gamma^{IJ}_{AB}\pi^-_B+\eta^-_{\dot
A}\tilde\gamma^I_{\dot A\dot B}\pi^{+}_{\dot B})\approx0
\end{array}
\end{equation}
complementing those that define the D.B. Constraints
(\ref{m+2-2})-(\ref{mij}) also include the
contributions of other supertwistor components and their conjugate
momenta and coincide with the linear combinations of primary
constraints (\ref{mumom})-(\ref{hmominverse})
\begin{equation}
\begin{array}{c}
\hat M^{+2-2}(\sigma)=v^+_{\alpha A}T^{\alpha-}_A-v^-_{\alpha\dot
A}T^{\alpha+}_{\dot A}-v^{\alpha-}_Ap^+_{(v)\alpha A}
+v^{\alpha+}_{\dot A}p^-_{(v)\alpha\dot
A}+\mu^{\alpha+}_A\Phi^-_{\alpha
A}-\mu^{\alpha-}_{\dot A}\Phi^+_{\alpha\dot A}\\[0.2cm]
+\eta^+_AD^-_A-\eta^-_{\dot A}D^+_{\dot A}\approx0,\\[0.2cm]
\hat M^{+2I}(\sigma)=-v^+_{\alpha A}\gamma^I_{A\dot
A}T^{\alpha+}_{\dot A}+v^{\alpha+}_{\dot A}\tilde\gamma^I_{\dot
AA}p^+_{(v)\alpha
A}-\mu^{\alpha+}_A\gamma^I_{A\dot A}\Phi^+_{\alpha\dot A}-\eta^+_A\gamma^I_{A\dot A}D^+_{\dot A}\approx0,\\[0.2cm]
\hat M^{-2I}(\sigma)=-v^-_{\alpha\dot A}\tilde\gamma^I_{\dot
AA}T^{\alpha-}_{A}+v^{\alpha-}_{A}\gamma^I_{A\dot
A}p^-_{(v)\alpha\dot A}
-\mu^{\alpha-}_{\dot A}\tilde\gamma^I_{\dot AA}\Phi^-_{\alpha A}-\eta^-_{\dot A}\tilde\gamma^I_{\dot AA}D^-_A\approx0,\\[0.2cm]
\hat M^{IJ}(\sigma)=-\frac12(v^+_{\alpha
A}\gamma^{IJ}_{AB}T^{\alpha-}_B+v^-_{\alpha\dot
A}\tilde\gamma^{IJ}_{\dot A\dot B}T^{\alpha+}_{\dot B}
+v^{\alpha-}_A\gamma^{IJ}_{AB}p^+_{(v)\alpha B}+v^{\alpha+}_{\dot
A}\tilde\gamma^{IJ}_{\dot A\dot B}p^-_{(v)\alpha\dot
B}\\[0.2cm]
+\mu^{\alpha+}_A\gamma^{IJ}_{AB}\Phi^-_{\alpha
B}+\mu^{\alpha-}_{\dot A}\tilde\gamma^{IJ}_{\dot A\dot
B}\Phi^+_{\alpha \dot
B}+\eta^+_A\gamma^{IJ}_{AB}D^-_B+\eta^-_{\dot
A}\tilde\gamma^I_{\dot A\dot B}D^{+}_{\dot B})\approx0
\end{array}
\end{equation}
modulo the twistor constraints (\ref{x3}), (\ref{x5}). These so
called covariant momentum densities (\ref{m+2-2})-(\ref{mij})
satisfy on P.B. the relations of $so(1,9)$ algebra and are the
generators of infinitesimal local $SO(1,9)$ transformations acting
on the supertwistor variables.

Since the twistor formulation is characterized by the presence of twistor
constraints (\ref{x3}), (\ref{x5}) it is helpful to introduce D.B. that take
them into account as well. Considering the projections of constraints (\ref{mumom})
\begin{equation}\label{phidb}
\begin{array}{c}
\Phi^{+2}_{\dot A\dot B}(\sigma)=v^{\alpha+}_{\dot A}\Phi^+_{\alpha\dot B}-(\dot A\leftrightarrow\dot B)\approx0,\
\Phi^{-2}_{AB}(\sigma)=v^{\alpha-}_{A}\Phi^-_{\alpha B}-(A\leftrightarrow
B)\approx0,\\[0.2cm]
\Phi_{A\dot B}(\sigma)=v^{\alpha-}_A\Phi^+_{\alpha\dot
B}-v^{\alpha+}_{\dot B}\Phi^-_{\alpha A}\approx0,\\[0.2cm]
\Phi^{m_1...m_5}(\sigma)=v^+_{\alpha A}\tilde\sigma^{m_1...m_5\alpha\beta}\Phi^-_{\beta A}+v^-_{\alpha\dot A}\tilde\sigma^{m_1...m_5\alpha\beta}\Phi^+_{\alpha\dot A}\approx0
\end{array}
\end{equation}
conjugate to the twistor constraints (\ref{x3}), (\ref{x5}) on the P.B.
\begin{equation}
\{\Phi^{+2}_{\dot A\dot B}(\sigma),N^{-2}_{\dot C\dot D}(\sigma')\}=2(\delta_{\dot A\dot D}\delta_{\dot B\dot C}-\delta_{\dot A\dot C}\delta_{\dot B\dot D})\delta(\sigma-\sigma'),
\end{equation}
\begin{equation}
\{\Phi^{-2}_{AB}(\sigma),N^{+2}_{CD}(\sigma')\}=2(\delta_{AD}\delta_{BC}-\delta_{AC}\delta_{BD})\delta(\sigma-\sigma'),
\end{equation}
\begin{equation}
\{\Phi_{A\dot B}(\sigma),N_{C\dot D}(\sigma')\}=-2\delta_{AC}\delta_{\dot B\dot D}\delta(\sigma-\sigma'),
\end{equation}
\begin{equation}
\{\Phi^{m_1...m_5}(\sigma),N^{n_1...n_5}(\sigma')\}=2(\tilde\sigma^{m_1...m_5}\sigma^{n_1...n_5})\delta(\sigma-\sigma'),
\end{equation}
with all other P.B. vanishing in the strong sense, we can introduce the second stage or twistor D.B. that in
the subspace of the phase-space defined by the projections of constraints (\ref{mumom})
\begin{equation}
\begin{array}{c}
\Phi^{+2}(\sigma)=2v^{\alpha+}_{\dot A}p^+_{(\mu)\alpha\dot
A}+\frac{1}{\alpha'}(\rho^{\tau+2}-is\varphi^{+2}_\sigma)\approx0,\\[0.2cm]
\Phi^{-2}(\sigma)=2v^{\alpha-}_{A}p^-_{(\mu)\alpha
A}+\frac{1}{\alpha'}(\rho^{\tau-2}-is\varphi^{-2}_\sigma)\approx0,\\[0.2cm]
 \Phi^I(\sigma)=-v^{\alpha-}_A\gamma^I_{A\dot A}p^+_{(\mu)\alpha\dot A}-v^{\alpha+}_{\dot A}\tilde\gamma^I_{\dot AA}p^-_{(\mu)\alpha A}
 -\frac{is}{\alpha'}\varphi^I_\sigma\approx0,
 \end{array}
\end{equation}
that complement those of (\ref{phidb}) as well as $so(1,9)$
generators (\ref{m+2-2})-(\ref{mij}) and primary constraints
(\ref{etamom}), (\ref{rhotau}), (\ref{rhosigma}) coincide with the
P.B.

So we arrive at the following expression for the total Hamiltonian
density
\begin{equation}\label{h-dense}
\begin{array}{rl}
H_t(\tau,\sigma)=&\frac{\rho^{\sigma+2}}{2\alpha'}(\omega^{-2}_\sigma+\rho^{\tau-2})+\frac{\rho^{\sigma-2}}{2\alpha'}(\omega^{+2}_\sigma-\rho^{\tau+2})
+a^{+2}\Phi^{-2}+a^{-2}\Phi^{+2}+a^I\Phi^I\\[0.2cm]
+&l^{+2-2}M^{+2-2}+l^{+2I}M^{-2I}+l^{-2I}M^{+2I}+l^{IJ}M^{IJ}\\[0.2cm]
+&b^{\mu+2}P^{-2}_\mu+b^{\mu-2}P^{+2}_\mu+\xi^+_AD^-_A+\xi^-_{\dot A}D^+_{\dot A}
\end{array}
\end{equation}
as the linear combination of the remaining primary constraints
with the bosonic $a(\sigma)$, $b(\sigma)$, $l(\sigma)$ and
fermionic $\xi(\sigma)$ Lagrange multipliers to be determined from
the consistency requirement. In the canonical
formalism evolution of any function of the phase-space variables
is defined by its P.B. with the Hamiltonian
\begin{equation}
\dot f(Q,P)=\{f,\mathcal H\},
\end{equation}
where $\mathcal H=\int d\sigma H_t(\tau,\sigma)$.
Following the Dirac method the consistency requirement for the constraints is that they are weakly conserved, i.e. their P.B. with the Hamiltonian $\mathcal H$ weakly vanish\footnote{P.B. of the primary and secondary constraints are given in Appendix A.}. In the case under consideration we find that the conservation of constraints $P^{\pm2}_\sigma\approx0$ yields the pair of secondary ones
\begin{equation}\label{secons}
\omega^{\pm2}_\sigma\mp\rho^{\tau\pm2}\approx0,
\end{equation}
whereas from the conservation conditions for
$P^{\pm2}_\tau\approx0$ we obtain the following equations for the
Lagrange multipliers
\begin{equation}\label{a-longit}
a^{+2}=-{\textstyle\frac12}\rho^{\sigma+2}+{\textstyle\frac{i}{16}}\xi^+_A\eta^+_A,\quad
a^{-2}={\textstyle\frac12}\rho^{\sigma-2}+{\textstyle\frac{i}{16}}\xi^-_{\dot
A}\eta^-_{\dot A}.
\end{equation}
P.B. of $M^{IJ}\approx0$ with $\mathcal H$ weakly vanish, while
of $M^{+2-2}\approx0$ with $\mathcal H$ weakly vanish
provided one uses (\ref{a-longit}). Conservation of the
$so(1,9)/(so(1,1)\times so(8))$ coset generators
$M^{\pm2I}\approx0$ results in the secondary constraints
\begin{equation}\label{secons2}
\omega^I_\sigma\approx0
\end{equation}
and the restriction for Lagrange multipliers $a^I$
\begin{equation}\label{a-transv}
a^I={\textstyle\frac{i}{16}}(\eta^+\gamma^I\xi^--\xi^+\gamma^I\eta^-).
\end{equation}
Evaluating P.B. of $\Phi^{\pm2}\approx0$ and $\mathcal H$ one arrives at the equations
\begin{equation}\label{b+2}
b^{\tau+2}=-\partial_\sigma\rho^{\sigma+2}-{\textstyle\frac12}\Omega^{+2-2}_\sigma\rho^{\sigma+2}+2l^{+2-2}\rho^{\tau+2}+{\textstyle\frac{is}{4}}\mathcal D_\sigma\eta^+_{A}\xi^+_{A},
\end{equation}
\begin{equation}\label{b-2}
b^{\tau-2}=-\partial_\sigma\rho^{\sigma-2}+{\textstyle\frac12}\Omega^{+2-2}_\sigma\rho^{\sigma-2}-2l^{+2-2}\rho^{\tau-2}+{\textstyle\frac{is}{4}}\mathcal
D_\sigma\eta^-_{\dot A}\xi^-_{\dot A},
\end{equation}
while from the conservation conditions for $\Phi^I\approx0$ and $\omega^I_\sigma\approx0$ there stem the equations
\begin{equation}
\rho^{\tau-2}l^{+2I}+\rho^{\tau+2}l^{-2I}={\textstyle\frac12}(\rho^{\sigma-2}\Omega^{+2I}_\sigma+\rho^{\sigma+2}\Omega^{-2I}_\sigma)
-{\textstyle\frac{is}{8}}(\xi^+\gamma^I\mathcal D_\sigma\eta^--\mathcal D_\sigma\eta^+\gamma^I\xi^-),
\end{equation}
\begin{equation}
\rho^{\tau-2}l^{+2I}-\rho^{\tau+2}l^{-2I}={\textstyle\frac12}(\rho^{\sigma-2}\Omega^{+2I}_\sigma-\rho^{\sigma+2}\Omega^{-2I}_\sigma)
-{\textstyle\frac{i}{8}}(\xi^+\gamma^I\mathcal D_\sigma\eta^--\mathcal D_\sigma\eta^+\gamma^I\xi^-).
\end{equation}
Conservation of the secondary constraint $\omega^{+2}-\rho^{\tau+2}\approx0$ gives another equation for $b^{\tau+2}$
\begin{equation}
b^{\tau+2}=-\partial_\sigma\rho^{\sigma+2}-{\textstyle\frac12}\Omega^{+2-2}_\sigma\rho^{\sigma+2}+2l^{+2-2}\rho^{\tau+2}+{\textstyle\frac{i}{4}}\xi^+_A\mathcal D_\sigma\eta^+_{A}.
\end{equation}
Its comparison with (\ref{b+2}) reveals that either $s=-1$ or $\xi^+_A\sim\mathcal D_\sigma\eta^+_A$.
Analogously considering the conservation of the secondary constraint
$\omega^{-2}_\sigma+\rho^{\tau-2}\approx0$ we derive another
equation for $b^{\tau-2}$
\begin{equation}
b^{\tau-2}=-\partial_\sigma\rho^{\sigma-2}+{\textstyle\frac12}\Omega^{+2-2}_\sigma\rho^{\sigma-2}-2l^{+2-2}\rho^{\tau-2}+{\textstyle\frac{i}{4}}\mathcal D_\sigma\eta^-_{\dot A}\xi^-_{\dot A}.
\end{equation}
Its compatibility with (\ref{b-2}) requires either $s=1$ or
$\xi^-_{\dot A}\sim\mathcal D_\sigma\eta^-_{\dot A}$.
There remain to consider the consistency conditions for the fermionic
constraints (\ref{etamom}). For $D^-_A\approx0$ we obtain
$\xi^+_A=-\frac{\rho^{\sigma-2}}{\rho^{\tau-2}}\mathcal
D_\sigma\eta^+_A$ when $s=1$, while in the case $s=-1$ no new
restrictions on the Lagrange multipliers arise. The conservation
condition for $D^+_{\dot A}\approx0$ yields
$\xi^-_{\dot A}=-\frac{\rho^{\sigma+2}}{\rho^{\tau+2}}\mathcal
D_\sigma\eta^-_{\dot A}$ when $s=-1$, but is trivial when $s=1$.
So we conclude that for $s=1$
$\xi^+_A=-\frac{\rho^{\sigma-2}}{\rho^{\tau-2}}\mathcal
D_\sigma\eta^+_A$, while $\xi^-_{\dot A}$ remains undetermined,
whereas when $s=-1$ $\xi^-_{\dot
A}=-\frac{\rho^{\sigma+2}}{\rho^{\tau+2}}\mathcal
D_\sigma\eta^-_{\dot A}$, but $\xi^+_A$ is free.

Upon substitution of the above derived expressions for the
Lagrange multipliers back into the Hamiltonian density
(\ref{h-dense}) it turns into the following linear combination of the
first-class constraints
\begin{equation}
H_{t,s=1}=\rho^{\sigma+2}\Delta^{-2}_{(-)}+\rho^{\sigma-2}\widetilde\Delta^{+2}_{(+)}+l^{+2-2}\widetilde M^{+2-2}+l^{IJ}M^{IJ}+b^{\sigma+2}P^{-2}_\sigma+b^{\sigma-2}P^{+2}_\sigma+\xi^-_{\dot A}\widetilde D^+_{\dot A}\approx0,
\end{equation}
where
\begin{equation}\label{delta-2-}
\Delta^{-2}_{(-)}(\sigma)={\textstyle\frac{1}{2\alpha'}}(\omega^{-2}_\sigma+\rho^{\tau-2})-{\textstyle\frac{1}{2}}\Phi^{-2}+\partial_\sigma
P^{-2}_\tau-{\textstyle\frac12}\Omega^{+2-2}_\sigma
P^{-2}_\tau+{\textstyle\frac{1}{2\rho^{\tau+2}}}\Omega^{-2I}_\sigma
M^{+2I}\approx0,
\end{equation}
and
\begin{equation}\label{corepa+}
\widetilde\Delta^{+2}_{(+)}(\sigma)=\Delta^{+2}_{(+)}-{\textstyle\frac{1}{\rho^{\tau-2}}}\mathcal
D_\sigma\eta^+_A\widetilde D^-_A\approx0
\end{equation}
are the generators of the reparametrizations\footnote{In the above
expressions and in what follows lower $\pm$ indices in brackets of
$\Delta^{\pm2}_k$ indicate the sign of the $\Phi^{\pm2}\approx0$
constraint contribution and should not be confused with the
$SO(1,1)$ indices.}. In (\ref{corepa+}) we introduced the
following combinations of the primary and secondary constraints
\begin{equation}\label{delta+2+}
\Delta^{+2}_{(+)}(\sigma)={\textstyle\frac{1}{2\alpha'}}(\omega^{+2}_\sigma-\rho^{\tau+2})+{\textstyle\frac{1}{2}}\Phi^{+2}+\partial_\sigma
P^{+2}_\tau+{\textstyle\frac12}\Omega^{+2-2}_\sigma
P^{+2}_\tau+{\textstyle\frac{1}{2\rho^{\tau-2}}}\Omega^{+2I}_\sigma
M^{-2I}\approx0,
\end{equation}
\begin{equation}\label{fermi2}
\widetilde
D^-_A(\sigma)=D^-_A+{\textstyle\frac{i}{16}}\eta^+_A\Phi^{-2}-{\textstyle\frac{i}{16}}\gamma^I_{A\dot
A}\eta^-_{\dot
A}\Phi^I-{\textstyle\frac{i}{8\rho^{\tau-2}}}\gamma^I_{A\dot
A}\mathcal D_\sigma\eta^-_{\dot A}M^{-2I}\approx0.
\end{equation}
Constraints (\ref{delta+2+}) and (\ref{delta-2-}) are the particular cases corresponding to $k=1$ and $k=-1$ respectively of the
more general constraints
\begin{equation}\label{delta+2k}
\Delta^{+2}_{k}(\sigma)={\textstyle\frac{1}{2\alpha'}}(\omega^{+2}_\sigma-\rho^{\tau+2})+{\textstyle\frac{k}{2}}\Phi^{+2}+\partial_\sigma
P^{+2}_\tau+{\textstyle\frac12}\Omega^{+2-2}_\sigma
P^{+2}_\tau+{\textstyle\frac{1}{2\rho^{\tau-2}}}\Omega^{+2I}_\sigma
M^{-2I}\approx0,
\end{equation}
and
\begin{equation}\label{delta-2k}
\Delta^{-2}_{k}(\sigma)={\textstyle\frac{1}{2\alpha'}}(\omega^{-2}_\sigma+\rho^{\tau-2})+{\textstyle\frac{k}{2}}\Phi^{-2}+\partial_\sigma
P^{-2}_\tau-{\textstyle\frac12}\Omega^{+2-2}_\sigma
P^{-2}_\tau+{\textstyle\frac{1}{2\rho^{\tau+2}}}\Omega^{-2I}_\sigma
M^{+2I}\approx0
\end{equation}
to be used below. Other bosonic first class-constraints
\begin{equation}\label{M+2-2}
\widetilde M^{+2-2}=M^{+2-2}+2\rho^{\tau+2}P^{-2}_\tau-2\rho^{\tau-2}P^{+2}_\tau\approx0
\end{equation}
and (\ref{mij}) generate $SO(1,1)\times SO(8)$ gauge transformations.
8 fermionic first-class constraints
\begin{equation}\label{conskappa}
\widetilde D^+_{\dot A}=D^+_{\dot A}+{\textstyle\frac{i}{16}}\eta^-_{\dot A}\Phi^{+2}-{\textstyle\frac{i}{16}}\tilde\gamma^I_{\dot AA}\eta^+_{A}\Phi^I-{\textstyle\frac{i}{8\rho^{\tau-2}}}\tilde\gamma^I_{\dot AA}\mathcal D_\sigma\eta^+_{A}M^{-2I}-{\textstyle\frac{i}{4}}\mathcal D_\sigma\eta^-_{\dot A}P^{+2}_\tau\approx0
\end{equation}
are responsible for the $\kappa-$symmetry.

When $s=-1$ we get
\begin{equation}
H_{t,s=-1}=\rho^{\sigma+2}\widetilde\Delta^{-2}_{(-)}+\rho^{\sigma-2}\Delta^{+2}_{(+)}+l^{+2-2}\widetilde
M^{+2-2}+l^{IJ}M^{IJ}+b^{\sigma+2}P^{-2}_\sigma+b^{\sigma-2}P^{+2}_\sigma+\xi^+_{A}\widetilde
D'^-_{A}\approx0,
\end{equation}
where
\begin{equation}\label{wwd--}
\widetilde\Delta^{-2}_{(-)}=\Delta^{-2}_{(-)}-{\textstyle\frac{1}{\rho^{\tau+2}}}\mathcal
D_\sigma\eta^-_{\dot A}\widetilde D'^+_{\dot
A}\approx0.
\end{equation}
In (\ref{wwd--}) the second-class constraints
$\widetilde D'^+_{\dot A}\approx0$ are defined as
\begin{equation}
\widetilde D'^+_{\dot A}=D^+_{\dot
A}+{\textstyle\frac{i}{16}}\eta^-_{\dot
A}\Phi^{+2}-{\textstyle\frac{i}{16}}\tilde\gamma^I_{\dot
AA}\eta^+_{A}\Phi^I+{\textstyle\frac{i}{8\rho^{\tau+2}}}\tilde\gamma^I_{\dot
AA}\mathcal D_\sigma\eta^+_{A}M^{+2I}\approx0,
\end{equation}
whereas the generators of the $\kappa-$symmetry equal
\begin{equation}
\widetilde
D'^-_A(\sigma)=D^-_A+{\textstyle\frac{i}{16}}\eta^+_A\Phi^{-2}-{\textstyle\frac{i}{16}}\gamma^I_{A\dot
A}\eta^-_{\dot
A}\Phi^I+{\textstyle\frac{i}{8\rho^{\tau+2}}}\gamma^I_{A\dot
A}\mathcal D_\sigma\eta^-_{\dot A}M^{+2I}+{\textstyle\frac{i}{4}}\mathcal
D_\sigma\eta^+_AP^{-2}_\tau\approx0.
\end{equation}
In what follows for definiteness we concentrate on exploring the $s=1$ case.

Now consider the canonical form of the irreducible
$\kappa-$symmetry transformations generated on P.B. by the
fermionic first-class constraints (\ref{conskappa}) according to
the rule
\begin{equation}
\delta_\kappa f(\tau,\sigma)=\left\{{\textstyle\int}
d\sigma'\kappa^-_{\dot B}(\sigma')\widetilde
D^+_{\dot B}(\sigma'),f\right\}.
\end{equation}
Straightforward calculation yields that the supertwistor components transform as
\begin{equation}
\delta_\kappa\mu^{\alpha-}_{\dot A}={\textstyle\frac{i}{2}}(\kappa^-_{\dot A}\eta^+_A+{\textstyle\frac18}(\kappa^-\tilde\gamma^I\eta^+)\tilde\gamma^I_{\dot AA})v^{\alpha-}_A+
{\textstyle\frac{i}{2}}(\kappa^-_{\dot A}\eta^-_{\dot B}+\eta^-_{\dot A}\kappa^-_{\dot B}+{\textstyle\frac14}\delta_{\dot A\dot B}(\kappa^-\eta^-))v^{\alpha+}_{\dot B},
\end{equation}
\begin{equation}
\delta_\kappa v^-_{\alpha\dot A}=0,\quad\delta_\kappa\eta^-_{\dot
A}=\kappa^-_{\dot A},
\end{equation}
\begin{equation}
\delta_\kappa\mu^{\alpha+}_A={\textstyle\frac{i}{8\rho^{\tau-2}}}(\kappa^-\tilde\gamma^I\mathcal D_\sigma\eta^+)\gamma^I_{A\dot A}\mu^{\alpha-}_{\dot A}
+{\textstyle\frac{i}{16}}(\kappa^-\tilde\gamma^I\eta^+)\gamma^I_{A\dot A}v^{\alpha+}_{\dot A}+{\textstyle\frac{i}{2}}\eta^+_A\kappa^-_{\dot B}v^{\alpha+}_{\dot B},
\end{equation}
\begin{equation}
\delta_\kappa v^+_{\alpha A}(\eta^+_A)={\textstyle\frac{i}{8\rho^{\tau-2}}}(\kappa^-\tilde\gamma^I\mathcal D_\sigma\eta^+)\gamma^I_{A\dot A}v^-_{\alpha\dot A}(\eta^-_{\dot A}).
\end{equation}
 We note that the $2d$ covariant form of the $\kappa-$symmetry
transformations derived in the framework of the Lagrangian approach \cite{CQG07} reduces to the
above expressions provided one replaces all the
$\tau$-derivatives of the coordinates using their equations of motion.

\section{The second-class constraints and Dirac brackets}

The twistor realization of the 33
bosonic and 8 fermionic first-class constraints by which the $D=10$ superstring in the twistor formulation is characterized has been exhibited above. The remaining
primary and secondary constraints are of the second class. They
can be classified according to their grading and the $SO(8)$
representation. 4 vector constraints are represented by
(\ref{m+2i}), (\ref{m-2i}) and
\begin{equation}\label{deltai}
\Delta^I_{k}(\sigma)={\textstyle\frac{1}{\alpha'}}\omega^I_\sigma+k\Phi^I-\Omega^{+2I}_\sigma
P^{-2}_\tau-\Omega^{-2I}_\sigma P^{+2}_\tau -{\textstyle\frac12}\mathscr
D^{IJ}_\sigma\left({\textstyle\frac{M^{+2J}}{\rho^{\tau+2}}}\right)-{\textstyle\frac12}\mathscr
D^{IJ}_\sigma\left({\textstyle\frac{M^{-2J}}{\rho^{\tau-2}}}\right)\approx0,
\end{equation}
where $\mathscr
D^{IJ}_\sigma=\delta^{IJ}\partial_\sigma-\Omega^{IJ}_\sigma$ is
the world-sheet projected $so(8)$ covariant differential. 4 scalar
constraints can be chosen as (\ref{rhotau}) and
$\Delta^{+2}_{(-)}\approx0$,
$\Delta^{-2}_{(+)}\approx0$ defined in (\ref{delta+2k}),
(\ref{delta-2k}). Finally there are 8 fermionic second-class
constraints (\ref{fermi2}).

In the canonical approach one of the possible options to take into
account the second-class constraints is to introduce D.B.
\begin{equation}\label{db}
\{f(\sigma),g(\sigma')\}_{D.B.}=\{f(\sigma),g(\sigma')\}-\{f(\sigma),\chi_{\mathfrak
m}\}\mathbf C^{-1\mathfrak{mn}}\{\chi_{\mathfrak n},g(\sigma')\},
\end{equation}
where $\chi_{\mathfrak m}$ denotes the set of the second-class constraints and $\mathbf C^{-1\mathfrak{mn}}$ is inverse to the Dirac matrix
\begin{equation}
\mathbf C_{\mathfrak{mn}}(\sigma,\sigma')=\{\chi_{\mathfrak
m}(\sigma),\chi_{\mathfrak n}(\sigma')\}. \end{equation}
For the
above choice of the second-class constraints set the Dirac matrix
acquires the form
\begin{equation}\label{dirm}
\mathbf C_{\mathfrak{mn}}=J_{\mathfrak{mn}}+\Lambda_{\mathfrak{mn}},
\end{equation}
where $J_{\mathfrak{mn}}$ is the block-diagonal graded antisymmetric
matrix and $\Lambda_{\mathfrak{mn}}$ depends linearly on the
constraints\footnote{Expressions for the P.B. of the second-class
constraints are given in Appendix B.}. Explicitly
$J_{\mathfrak{mn}}$ reads
\begin{equation}
\alpha'J=
\begin{array} {l|ccccccccc|}
 \multicolumn{1}{c}{}
&{\scriptstyle M^{+2J}}&{\scriptstyle\Delta^J_{(-)}}&{\scriptstyle
M^{-2J}} &{\scriptstyle\Delta^J_{(+)}} &
{\scriptstyle\Delta^{+2}_{(-)}} & {\scriptstyle
P^{-2}_\tau} & {\scriptstyle\Delta^{-2}_{(+)}} &
{\scriptstyle P^{+2}_\tau} &
\multicolumn{1}{c}{\scriptstyle\widetilde D^-_B}\\[0.2cm]
\cline{2-10}
{\scriptstyle M^{+2I}} &{\scriptstyle0}&\multicolumn{1}{c|}{\scriptstyle-2\rho^{\tau+2}\delta^{IJ}}&&&&&&&\\[0.2cm]
{\scriptstyle\Delta^I_{(-)}} &{\scriptstyle2\rho^{\tau+2}\delta^{IJ}}&\multicolumn{1}{c|}{\scriptstyle0}&&&&&&&\\[0.2cm] \cline{2-5}
{\scriptstyle M^{-2I}} &&&\multicolumn{1}{|c}{\scriptstyle0}&\multicolumn{1}{c|}{\scriptstyle2\rho^{\tau-2}\delta^{IJ}}&&&&&\\[0.2cm]
{\scriptstyle\Delta^I_{(+)}} &&&\multicolumn{1}{|c}{\scriptstyle-2\rho^{\tau-2}\delta^{IJ}}&\multicolumn{1}{c|}{\scriptstyle0}&&&&{\mathbf 0}&\\[0.2cm] \cline{4-7}
{\scriptstyle\Delta^{+2}_{(-)}} &&&&&\multicolumn{1}{|c}{\scriptstyle0}&\multicolumn{1}{c|}{\scriptstyle1}&&&\\[0.2cm]
{\scriptstyle P^{-2}_\tau} &&&&&\multicolumn{1}{|c}{\scriptstyle-1}&\multicolumn{1}{c|}{\scriptstyle0}&&&\\[0.2cm] \cline{6-9}
{\scriptstyle\Delta^{-2}_{(+)}} &&& {\mathbf 0}&&&&\multicolumn{1}{|c}{\scriptstyle0}&\multicolumn{1}{c|}{\scriptstyle-1}&\\[0.2cm]
{\scriptstyle P^{+2}_\tau}  &&&&&&&\multicolumn{1}{|c}{\scriptstyle1}&\multicolumn{1}{c|}{\scriptstyle0}&\\[0.2cm]
\cline{8-10}
{\scriptstyle\widetilde D^-_A} &&&&&&&&&\multicolumn{1}{|c|}{\scriptstyle\frac{i\rho^{\tau-2}}{4}\delta_{AB}}\\
\cline{2-10}
\end{array}\ \delta(\sigma-\sigma').
\end{equation}
Then the inverse to the Dirac matrix (\ref{dirm}) is given by
\begin{equation}
\mathbf C^{-1}=(I+J^{-1}\Lambda)J^{-1}
\end{equation}
and can be expanded as the series
\begin{equation}\label{c-1}
\mathbf C^{-1}=J^{-1}-J^{-1}\Lambda\ J^{-1}+ J^{-1}\Lambda\
J^{-1}\Lambda\ J^{-1}-J^{-1}\Lambda\ J^{-1}\Lambda\ J^{-1}\Lambda
J^{-1}+...
\end{equation}
Since $J$ is proportional to $(\alpha')^{-1}$ and its inverse
depends on $\alpha'$ the inverse Dirac matrix is presented as the
series in $\alpha'$ \footnote{Note, however, that some entries of
$\Lambda$ have implicit dependence on $(\alpha')^{-1}$ through constraints (\ref{delta+2k}), (\ref{delta-2k}), (\ref{deltai}).}.
Expansion (\ref{c-1}) suggests that we can evaluate $\mathbf
C^{-1}$ perturbatively with the leading contribution determined
by $J^{-1}$ and the D.B. acquire the form
\begin{equation}\label{db-leading}
\begin{array}{c}
\{f(\sigma),g(\sigma')\}_{D.B.}=\{f(\sigma),g(\sigma')\}-4i\alpha'\int\frac{d\sigma''}{\rho^{\tau-2}(\sigma'')}\{f(\sigma),\widetilde D^-_A(\sigma'')\}\{\widetilde D^-_A(\sigma''),g(\sigma')\}\\[0.2cm]
-\frac{\alpha'}{2}\int\frac{d\sigma''}{\rho^{\tau+2}(\sigma'')}(\{f(\sigma),M^{+2I}(\sigma'')\}\{\Delta^I_{(-)}(\sigma''),g(\sigma')\}-\{f(\sigma),\Delta^I_{(-)}(\sigma'')\}\{M^{+2I}(\sigma''),g(\sigma')\})\\[0.2cm]
+\frac{\alpha'}{2}\int\frac{d\sigma''}{\rho^{\tau-2}(\sigma'')}(\{f(\sigma),M^{-2I}(\sigma'')\}\{\Delta^I_{(+)}(\sigma''),g(\sigma')\}-\{f(\sigma),\Delta^I_{(+)}(\sigma'')\}\{M^{-2I}(\sigma''),g(\sigma')\})\\[0.2cm]
+\alpha'\int d\sigma''(\{f(\sigma),\Delta^{+2}_{(-)}(\sigma'')\}\{P^{-2}_\tau(\sigma''),g(\sigma')\}-\{f(\sigma),P^{-2}_\tau(\sigma'')\}\{\Delta^{+2}_{(-)}(\sigma''),g(\sigma')\})\\[0.2cm]
-\alpha'\int
d\sigma''(\{f(\sigma),\Delta^{-2}_{(+)}(\sigma'')\}\{P^{+2}_\tau(\sigma''),g(\sigma')\}-
\{f(\sigma),P^{+2}_\tau(\sigma'')\}\{\Delta^{-2}_{(+)}(\sigma''),g(\sigma')\})+O(J^{-2})
\end{array}
\end{equation}

Expressions (\ref{db}) and (\ref{db-leading}) can be used to
evaluate the D.B. relations of the superstring phase-space
variables. So for the two sets of supertwistors
(\ref{twistor10str}) associated with the world-sheet light-like
directions we get
\begin{equation}
\begin{array}{c}
\{Z^{\Lambda+}_A(\sigma),Z^{\Sigma+}_B(\sigma')\}_{D.B.}={\textstyle\frac{4i\alpha'}{\rho^{\tau-2}}}D^{\Lambda}_{AC}D^{\Sigma}_{BC}\delta(\sigma-\sigma')\\[0.2cm]
+{\textstyle\frac{\alpha'}{2\rho^{\tau-2}}}\gamma^I_{A\dot
A}\gamma^{I}_{B\dot B}(V^{\Lambda+}_{\dot A}Z^{\Sigma-}_{\dot
B}-Z^{\Lambda-}_{\dot A}V^{\Sigma+}_{\dot
B})\delta(\sigma-\sigma')\\[0.2cm]
+{\textstyle\frac{\alpha'}{2\rho^{\tau-2}}}\gamma^I_{A\dot
A}Z^{\Lambda-}_{\dot A}(\sigma)\gamma^{J}_{B\dot B}\mathscr
D^{IJ}_\sigma\left(\frac{1}{\rho^{\tau-2}}Z^{\Sigma-}_{\dot
B}(\sigma)\delta(\sigma-\sigma')\right)+O(\alpha'^2),
\end{array}
\end{equation}
\begin{equation}
\begin{array}{c}
\{Z^{\Lambda-}_{\dot A}(\sigma),Z^{\Sigma-}_{\dot
B}(\sigma')\}_{D.B.}=
{\textstyle\frac{4i\alpha'}{\rho^{\tau-2}}}D^{\Lambda-2}_{\dot
AC}D^{\Sigma-2}_{\dot BC}\delta(\sigma-\sigma')\\[0.2cm]
+ {\textstyle\frac{\alpha'}{2\rho^{\tau+2}}}\tilde\gamma^I_{\dot
AA}\tilde\gamma^I_{\dot
BB}(V^{\Lambda-}_AZ^{\Sigma+}_B-Z^{\Lambda+}_AV^{\Sigma-}_B)
\delta(\sigma-\sigma')\\[0.2cm]
-{\textstyle\frac{\alpha'}{2\rho^{\tau+2}}}\tilde\gamma^I_{\dot
AA}Z^{\Lambda+}_{A}(\sigma)\tilde\gamma^{J}_{\dot BB}\mathscr
D^{IJ}_\sigma\left(\frac{1}{\rho^{\tau+2}}Z^{\Sigma+}_{B}(\sigma)\delta(\sigma-\sigma')\right)+O(\alpha'^2),
\end{array}
\end{equation}
as well as
\begin{equation}
\begin{array}{c}
\{Z^{\Lambda+}_A(\sigma),Z^{\Sigma-}_{\dot
B}(\sigma')\}_{D.B.}=\frac{4i\alpha'}{\rho^{\tau-2}}D^{\Lambda}_{AC}D^{\Sigma-2}_{\dot
BC}\delta(\sigma-\sigma')\\[0.2cm]
+\frac{\alpha'}{2}\gamma^I_{A\dot A}\tilde\gamma^I_{\dot
BB}(\frac{1}{\rho^{\tau+2}}V^{\Lambda+}_{\dot
A}Z^{\Sigma+}_B-\frac{1}{\rho^{\tau-2}}Z^{\Lambda-}_{\dot
A}V^{\Sigma-}_B)\delta(\sigma-\sigma')+O(\alpha'^2),
\end{array}
\end{equation}
where the inverse spinor harmonic matrix components have been
promoted to supertwistors $V^{\Lambda-}_A=(v^{\alpha-}_A,\ 0,\
0)$, $V^{\Lambda+}_{\dot A}=(v^{\alpha+}_{\dot A},\ 0,\ 0)$ and
the following quantities have been introduced
\begin{equation}
\begin{array}{c}
\{\tilde
D^-_A(\sigma),Z^{\Lambda+}_B(\sigma')\}=D^{\Lambda}_{AB}\delta(\sigma-\sigma'),\\[0.2cm]
D^{\Lambda}_{AB}=-D^{\Lambda}_{BA}=\frac{i}{8\rho^{\tau-2}}\gamma^I_{A\dot
A}\mathcal D_\sigma\eta^-_{\dot A}\gamma^I_{B\dot
B}Z^{\Lambda-}_{\dot B}+\frac{i}{2}(\delta_{AB}\eta^-_{\dot
B}+\frac18\gamma^I_{A\dot A}\eta^-_{\dot A}\gamma^I_{B\dot
B})V^{\Lambda+}_{\dot
B}\\[0.2cm]
+\frac{i}{2}(\delta_{AB}\eta^+_{C}-\delta_{AC}\eta^+_B+\frac18\delta_{BC}\eta^+_A)V^{\Lambda-}_C+\delta_{AB}J^{\Lambda},\quad
J^{\Lambda}=(0,\ 0,\ 1),
\end{array}
\end{equation}
\begin{equation}
\{\tilde D^-_A(\sigma),Z^{\Lambda-}_{\dot
B}(\sigma')\}=D^{\Lambda-2}_{A\dot B}\delta(\sigma-\sigma'),\
D^{\Lambda-2}_{A\dot B}=-D^{\Lambda-2}_{\dot
BA}=-{\textstyle\frac{i}{2}}(\delta_{AB}\eta^-_{\dot B}-
{\textstyle\frac18}\gamma^I_{A\dot A}\eta^-_{\dot
A}\tilde\gamma^I_{\dot BB})V^{\Lambda-}_{B}.
\end{equation}
To the first order in $\alpha'$ supertwistors also satisfy nonzero
D.B. relations with the world-sheet densities
\begin{equation}
\begin{array}{c}
\{Z^{\Lambda+}_A(\sigma),\rho^{\tau-2}(\sigma')\}_{D.B.}=\alpha'V^{\Lambda-}_A\delta(\sigma-\sigma')-\frac{\alpha'}{2\rho^{\tau-2}}\Omega^{-2I}_\sigma\gamma^I_{A\dot
A}Z^{\Lambda-}_{\dot A}\delta(\sigma-\sigma')+O(\alpha'^2),\\[0.2cm]
\{Z^{\Lambda-}_{\dot
A}(\sigma),\rho^{\tau+2}(\sigma')\}_{D.B.}=\alpha'V^{\Lambda+}_{\dot
A}\delta(\sigma-\sigma')+\frac{\alpha'}{2\rho^{\tau+2}}\Omega^{+2I}_\sigma\tilde\gamma^I_{\dot
AA}Z^{\Lambda+}_A\delta(\sigma-\sigma')+O(\alpha'^2)
\end{array}\end{equation}

D.B. deformation of P.B. relations for the phase-space variables results in the deformation of the
first-class constraint algebra. For the $\kappa-$symmetry
generators (\ref{conskappa}) one obtains
\begin{equation}
\begin{array}{rl}
\{\widetilde D^+_{\dot A}(\sigma),\widetilde D^+_{\dot
B}(\sigma')\}_{D.B.}=& \frac{i}{4}\delta_{\dot A\dot
B}\widetilde\Delta^{+2}_{(+)}\delta(\sigma-\sigma')\\[0.2cm]
-&\frac{\alpha'}{16\rho^{\tau+2}(\rho^{\tau-2})^2}\tilde\gamma^{I}_{\dot
AA}\mathcal D_\sigma\eta^+_A\tilde\gamma^{J}_{\dot BB}\mathcal
D_\sigma\eta^+_BM^{IJ}\Delta^{-2}_{(-)}\delta(\sigma-\sigma')\\[0.2cm]
+&\frac{\alpha'}{2}\Gamma^{+I}_{\dot
A}(\sigma)\mathscr D^{IJ}_\sigma\Gamma^{+J}_{\dot
B}(\sigma)\delta(\sigma-\sigma')+O(J^{-2}),
\end{array}
\end{equation}
where $\Gamma^{+I}_{\dot
A}=-\frac{1}{4\rho^{\tau+2}\rho^{\tau-2}}\tilde\gamma^{J}_{\dot
AA}\mathcal D_\sigma\eta^+_A(\frac12\delta^{JI}\widetilde
M^{+2-2}+M^{JI})$. The first term on the r.h.s. proportional to
the reparametrization generator (\ref{corepa+}) is the P.B. contribution, while
the terms containing products of $so(1,1)$, $so(8)$ generators and
the reparametrization generator (\ref{delta-2-}) correspond to the deformation.
The D.B. of the $\kappa-$symmetry and corresponding
reparametrization generators have the form
\begin{equation}
\begin{array}{rl}
\{\widetilde\Delta^{+2}_{(+)}(\sigma),\widetilde
D^+_{\dot
A}(\sigma')\}_{D.B.}=&\frac{i\alpha'}{4\rho^{\tau-2}}\tilde\gamma^I_{\dot
AA}\mathcal D_\sigma\eta^+_A \left(A^{+2I}-\frac12\mathcal
D_\sigma\eta^+\gamma^I\widetilde
D^+\right)\Delta^{-2}_{(-)}\delta(\sigma-\sigma')\\[0.2cm]
+&\frac{i\alpha'}{2}A^{+2I}(\sigma)\mathscr D^{IJ}_\sigma\Gamma^{+J}_{\dot A}(\sigma)\delta(\sigma-\sigma')+O(J^{-2}),
\end{array}
\end{equation}
where
\begin{equation}
A^{+2I}={\textstyle\frac{1}{\rho^{\tau+2}\rho^{\tau-2}}}\left[(\mathcal
D_\sigma\eta^+\gamma^I\widetilde D^+)\!+\!
\left(-{\textstyle\frac12}\delta^{IJ}\widetilde
M^{+2-2}+M^{IJ}\right)\!\!\left(\Omega^{+2J}_\sigma
+{\textstyle\frac{i}{4\rho^{\tau-2}}}(\mathcal
D_\sigma\eta^+\gamma^J\mathcal D_\sigma\eta^-)\right)\right].
\end{equation}
Observe that equal to zero P.B. contribution becomes supplemented by the terms quadratic in the first-class constraints. At the same time at the lowest order in
$J^{-1}$ D.B. of the $\kappa-$symmetry generators and the reparametrization generator (\ref{delta-2-}) coincide with the P.B.
\begin{equation}
\{\Delta^{-2}_{(-)}(\sigma),\widetilde
D^+_{\dot A}(\sigma')\}_{D.B.}={\textstyle\frac{i}{2}}\Gamma^{+I}_{\dot A}\Omega^{-2I}_\sigma\delta(\sigma-\sigma')+O(J^{-2})=-{\textstyle\frac{i}{8}}\tilde\gamma^I_{\dot
AA}\mathcal
D_\sigma\eta^+_AB^{-2I}\delta(\sigma-\sigma')+O(J^{-2}),
\end{equation}
where
$B^{-2I}=\frac{1}{\rho^{\tau+2}\rho^{\tau-2}}\left(\frac12\delta^{IJ}\widetilde
M^{+2-2}+M^{IJ}\right)\Omega^{-2J}_\sigma$. Equal to zero
diagonal P.B. relations of the reparametrization generators on
transition to D.B. receive contributions quadratic in $so(1,1)$
and $so(8)$
 generators
\begin{equation}
\{\widetilde\Delta^{+2}_{(+)}(\sigma),\widetilde\Delta^{+2}_{(+)}(\sigma')\}_{D.B.}=-{\textstyle\frac{\alpha'}{2}}A^{+2I}(\sigma)
\mathscr D^{IJ}_\sigma A^{+2J}(\sigma)\delta(\sigma-\sigma')+O(J^{-2}),
\end{equation}
\begin{equation}
\{\Delta^{-2}_{(-)}(\sigma),\Delta^{-2}_{(-)}(\sigma')\}_{D.B.}=
{\textstyle\frac{\alpha'}{2}}B^{-2I}(\sigma)\mathscr D^{IJ}_\sigma B^{-2J}(\sigma)\delta(\sigma-\sigma')+O(J^{-2}),
\end{equation}
while that of different reparametrization generators at the lowest
order in $J^{-1}$ become deformed by the terms proportional to the
product of the reparametrization generator (\ref{delta-2-}) with the
generators of the $\kappa-$symmetry and $so(1,1)$, $so(8)$
generators
\begin{equation}
\begin{array}{rl}
\{\widetilde\Delta^{+2}_{(+)}(\sigma),\Delta^{-2}_{(-)}(\sigma')\}_{D.B.}=&\frac{1}{2\rho^{\tau+2}\rho^{\tau-2}}
(\mathcal D_\sigma\eta^+\gamma^I\widetilde
D^+)\Omega^{-2I}_\sigma\left(1+\frac{\alpha'}{\rho^{\tau-2}}\Delta^{-2}_{(-)}\right)
\delta(\sigma-\sigma')\\[0.2cm]
-&\frac{1}{2}\left[
\Omega^{+2I}_\sigma\left(1+\frac{\alpha'}{\rho^{\tau-2}}\Delta^{-2}_{(-)}\right)+\frac{i}{4\rho^{\tau-2}}(\mathcal
D_\sigma\eta^+ \gamma^I\mathcal
D_\sigma\eta^-)\left(1+\frac{2\alpha'}{\rho^{\tau-2}}\Delta^{-2}_{(-)}\right)\right]\\[0.2cm]
\times &B^{-2I}\delta(\sigma-\sigma')+O(J^{-2}).
\end{array}
\end{equation}
P.B. of the $so(1,1)$ and $so(8)$ generators (\ref{M+2-2}), (\ref{mij}) and the second-class constraints are determined by their properties under the $SO(1,1)\times SO(8)$ transformations and hence are proportional to the second-class constraints. So that the D.B. involving $\widetilde M^{+2-2}\approx0$ and/or $M^{IJ}\approx0$ are equal to corresponding P.B.

\section{Conclusion}

The present paper investiges the canonical approach
application to the $D=10$ superstring first-order action involving
spinor harmonics and formulated in terms of supertwistor variables
generalizing Penrose-Ferber ones. We have identified the primary
and secondary constraints on the supertwistors and conjugate
momenta, analyzed their consistency and as a result obtained the
set of the first-class constraints that includes twistor
realizations of the reparametrization, $SO(1,1)\times SO(8)$ gauge
symmetry and $\kappa-$symmetry generators. The superstring model
is also characterized by the second-class
constraints that can be taken into account by constructing D.B. To
this end we have chosen the basis in the space of the second-class
constraints such that the Dirac matrix acquires the form of the
sum of block-diagonal graded antisymmetric matrix $J$ proportional
to $(\alpha')^{-1}$ and the one linear in the constraints. So the
D.B. can be evaluated perturbatively as the series in $J^{-1}$.
Introduction of D.B. leads to the deformation of the first-class
constraint algebra, the deformation parameter can be identified
with $\alpha'$. We have explicitly found the D.B. deformation of 
the first-class constraint algebra up to terms quadratic in the
constraints although it can be calculated to any order in
$J^{-1}$. One could expect some simplification of the expression
for D.B. by choosing such representation for the second-class
constraints for which the Dirac matrix becomes strongly equal to
$J$. This, however, requires addition to the obtained second-class
constraints of the terms containing higher powers of the
constraints to compensate weakly vanishing contributions to their
P.B. Another way to handle the second-class constraints is to bring
them by canonical transformation to the special form\footnote{Note
that the conjugate pairs of the second-class constraints
$(P^{\pm2}_\tau, \Delta^{\mp2}_{(\pm)})$ are already brought to
the special form.} and then solve with respect to the subset of
the canonically conjugate variables equal in number to the
second-class constraints \cite{Gitman}. Alternative mode could be
to consider the covariant supertwistor analogue of the semilightcone
gauge approach to the GS superstring quantization \cite{Carlip}.
Its examination for the superstring in the twistor
formulation has been initiated in \cite{IJMPA07} on the example of the 
$D=4$ model. All these possibilities are under consideration.

\section{Acknowledgements}

The author is obliged to A.A.~Zheltukhin for valuable discussions.

\appendix
\section{P.B. relations of the primary and secondary constraints}
Fermionic constraints (\ref{etamom}) satisfy the following nonzero P.B. relations between themselves
\begin{equation}
\begin{array}{rl}
\{D^-_A(\sigma),D^-_B(\sigma')\}=&{\textstyle\frac{1}{8\alpha'}}\delta_{AB}(is\omega^{-2}_\sigma-i\rho^{\tau-2}-s\varphi^{-2}_\sigma)\delta(\sigma-\sigma'),\\[0.2cm]
\{D^+_{\dot A}(\sigma),D^+_{\dot
B}(\sigma')\}=&{\textstyle\frac{1}{8\alpha'}}\delta_{\dot A\dot
B}(is\omega^{+2}_\sigma-i\rho^{\tau+2}-s\varphi^{+2}_\sigma)\delta(\sigma-\sigma'),\\[0.2cm]
\{D^-_A(\sigma),D^+_{\dot
B}(\sigma')\}=&\frac{s}{8\alpha'}\gamma^I_{A\dot
B}(\varphi^{I}_\sigma-i\omega^I_\sigma)\delta(\sigma-\sigma')
\end{array}
\end{equation}
and with the $\Phi$-constraints (\ref{mumom})
\begin{equation}
\begin{array}{c}
\{D^-_A(\sigma),\Phi^{+2}(\sigma')\}=\frac{is}{4\alpha'}\mathcal
D_\sigma\eta^+_A\delta(\sigma-\sigma'),\quad
\{D^-_A(\sigma),\Phi^{I}(\sigma')\}=\frac{is}{8\alpha'}\gamma^I_{A\dot
A}\mathcal D_\sigma\eta^-_{\dot A}\delta(\sigma-\sigma'),\\[0.2cm]
\{D^+_{\dot
A}(\sigma),\Phi^{-2}(\sigma')\}=\frac{is}{4\alpha'}\mathcal
D_\sigma\eta^-_{\dot A}\delta(\sigma-\sigma'),\quad \{D^+_{\dot
A}(\sigma),\Phi^{I}(\sigma')\}=\frac{is}{8\alpha'}\tilde\gamma^I_{\dot
AA}\mathcal D_\sigma\eta^+_{A}\delta(\sigma-\sigma').
\end{array}
\end{equation}
$\sigma-$components of the world-sheet projections of $\omega$ 1-forms that enter the secondary constraints (\ref{secons}), (\ref{secons2}) satisfy the P.B. relations with the fermionic constraints
\begin{equation}
\begin{array}{rl}
\{D^-_A(\sigma),\omega^{+2}_\sigma(\sigma')\}=&\frac{i}{8}\eta^+_A(\sigma)(\partial_\sigma-\frac12\Omega^{+2-2}_\sigma)\delta(\sigma-\sigma')+\frac{i}{4}(\mathcal D_\sigma\eta^+_A+\frac{1}{4}\gamma^I_{A\dot A}\eta^-_{\dot A}\Omega^{+2I}_\sigma)\delta(\sigma-\sigma'),\\[0.2cm]
\{D^-_A(\sigma),\omega^{-2}_\sigma(\sigma')\}=&\frac{i}{16}\gamma^I_{A\dot A}\eta^-_{\dot A}\Omega^{-2I}_\sigma\delta(\sigma-\sigma'),\\[0.2cm]
\{D^-_A(\sigma),\omega^{I}_\sigma(\sigma')\}=&-\frac{i}{16}\mathscr D^{IJ}_{\sigma'}\gamma^J_{A\dot
A}\eta^-_{\dot A}(\sigma)\delta(\sigma-\sigma')
+\frac{i}{8}(\gamma^I_{A\dot A}\mathcal D_\sigma\eta^-_{\dot
A}+\frac12\eta^+_A\Omega^{-2I}_\sigma)\delta(\sigma-\sigma'),\\[0.2cm]
\{D^+_{\dot A}(\sigma),\omega^{-2}_\sigma(\sigma')\}=&\frac{i}{8}\eta^-_{\dot A}(\sigma)(\partial_\sigma+\frac12\Omega^{+2-2}_\sigma)\delta(\sigma-\sigma')
+\frac{i}{4}(\mathcal D_\sigma\eta^-_{\dot A}+\frac{1}{4}\tilde\gamma^I_{\dot AA}\eta^+_{A}\Omega^{-2I}_\sigma)\delta(\sigma-\sigma'),\\[0.2cm]
\{D^+_{\dot A}(\sigma),\omega^{+2}(\sigma')\}=&\frac{i}{16}\tilde\gamma^I_{\dot AA}\eta^+_{A}\Omega^{+2I}_\sigma\delta(\sigma-\sigma'),\\[0.2cm]
\{D^+_{\dot A}(\sigma),\omega^{I}_\sigma(\sigma')\}=&-\frac{i}{16}\mathscr D^{IJ}_{\sigma'}\tilde\gamma^J_{\dot
AA}\eta^+_{A}(\sigma)\delta(\sigma-\sigma')
+\frac{i}{8}(\tilde\gamma^I_{\dot AA}\mathcal D_\sigma\eta^+_{A}+\frac12\eta^-_{\dot A}\Omega^{+2I}_\sigma)\delta(\sigma-\sigma'),
\end{array}
\end{equation}
and with the $\Phi$-constraints
\begin{equation}
\begin{array}{c}
\{\Phi^{\pm2}(\sigma),\omega^{\mp2}_\sigma(\sigma')\}=-2(\partial_\sigma\pm\frac12\Omega^{+2-2}_\sigma)\delta(\sigma-\sigma'),\\[0.2cm]
\{\Phi^I(\sigma),\omega^{\pm2}_\sigma(\sigma')\}=\{\omega^I_\sigma(\sigma),\Phi^{\pm2}(\sigma')\}=\Omega^{\pm2I}_\sigma\delta(\sigma-\sigma').
\end{array}
\end{equation}

The P.B. of $so(1,1)$ and $so(1,9)/(so(1,1)\times so(8))$ coset generators (\ref{m+2-2})-(\ref{m-2i}) with the fermionic constraints equal
\begin{equation}
\begin{array}{rl}
\{M^{+2-2}(\sigma),D^-_A(\sigma')\}=&-(D^-_A+\frac{i\rho^{\tau-2}}{8\alpha'}\eta^+_A)\delta(\sigma-\sigma'),\\[0.2cm]
\{M^{+2-2}(\sigma),D^+_{\dot A}(\sigma')\}=&(D^+_{\dot
A}+\frac{i\rho^{\tau+2}}{8\alpha'}\eta^-_{\dot
A})\delta(\sigma-\sigma'),\\[0.2cm]
\{M^{+2I}(\sigma),D^-_A(\sigma')\}=&\gamma^I_{A\dot A}(D^+_{\dot A}+\frac{i\rho^{\tau+2}}{16\alpha'}\eta^-_{\dot A})\delta(\sigma-\sigma'),\\[0.2cm]
\{M^{+2I}(\sigma),D^+_{\dot A}(\sigma')\}=&\frac{i\rho^{\tau+2}}{16\alpha'}\tilde\gamma^I_{\dot AA}\eta^+_{A}\delta(\sigma-\sigma'),\\[0.2cm]
\{M^{-2I}(\sigma),D^-_{A}(\sigma')\}=&\frac{i\rho^{\tau-2}}{16\alpha'}\gamma^I_{A\dot A}\eta^-_{\dot A}\delta(\sigma-\sigma'),\\[0.2cm]
\{M^{-2I}(\sigma),D^+_{\dot A}(\sigma')\}=&\tilde\gamma^I_{\dot
AA}(D^-_{A}+\frac{i\rho^{\tau-2}}{16\alpha'}\eta^+_{A})\delta(\sigma-\sigma')
\end{array}
\end{equation}
while with the $\Phi$-constraints read
\begin{equation}
\begin{array}{rl}
\{M^{+2-2}(\sigma),\Phi^{\pm2}(\sigma')\}=&\pm2(\Phi^{\pm2}-\frac{1}{\alpha'}\rho^{\tau\pm2})\delta(\sigma-\sigma'),\\[0.2cm]
\{M^{\pm2I}(\sigma),\Phi^{\mp2}(\sigma')\}=&-2\Phi^I\delta(\sigma-\sigma'),\\[0.2cm]
\{M^{\pm2I}(\sigma),\Phi^{J}(\sigma')\}=&-\delta^{IJ}(\Phi^{\pm2}-\frac{\rho^{\tau\pm2}}{\alpha'})\delta(\sigma-\sigma').
\end{array}
\end{equation}
Finally the P.B. of the $so(1,1)$ and $so(1,9)/(so(1,1)\times so(8))$ generators and the $\sigma-$components of $\omega$ 1-forms are given by
\begin{equation}
\begin{array}{rl}
\{M^{+2-2}(\sigma),\omega^{\pm2}_\sigma(\sigma')\}=&\pm2\omega^{\pm2}_\sigma\delta(\sigma-\sigma'),\\[0.2cm]
\{M^{\pm2I}(\sigma),\omega^{\mp2}_\sigma(\sigma')\}=&-2\omega^I_\sigma\delta(\sigma-\sigma'),\\[0.2cm]
\{M^{\pm2I}(\sigma),\omega^{J}_\sigma(\sigma')\}=&-\delta^{IJ}\omega^{\pm2}\delta(\sigma-\sigma').
\end{array}
\end{equation}
R.h.s. of the P.B. of $SO(8)$ generators $M^{IJ}\approx0$ with the fermionic constraints, $\Phi$-constraints and the $\sigma-$components of $\omega$ 1-forms are defined by their transformation properties under the $SO(8)$ transformations and exhibit no "anomalous" contributions as opposed to the above given P.B.

\section{P.B. relations of the second-class constraints}

P.B. of the $SO(8)$ vector constraints (\ref{m+2i}),
(\ref{m-2i}) and (\ref{deltai}) are given by
\begin{equation}
\begin{array}{rl}
\{\Delta^I_k(\sigma),\Delta^J_{k'}(\sigma')\}=&
\left(\frac{\Delta^{-2}_k}{\rho^{\tau-2}}(\sigma)+\frac{\Delta^{-2}_{k'}}{\rho^{\tau-2}}(\sigma')\right)\mathscr
D^{IJ}_\sigma\delta(\sigma-\sigma')
+\left(\frac{\Delta^{+2}_k}{\rho^{\tau+2}}(\sigma)+\frac{\Delta^{+2}_{k'}}{\rho^{\tau+2}}(\sigma')\right)\mathscr D^{IJ}_\sigma\delta(\sigma-\sigma')
\\[0.2cm]
+&\frac14\left(\frac{M^{+2K}}{\rho^{\tau+2}}-\frac{M^{-2K}}{\rho^{\tau-2}}\right)
\left(\frac{\Omega^{+2K}_\sigma}{\rho^{\tau+2}}-\frac{\Omega^{-2K}_\sigma}{\rho^{\tau-2}}\right)(\sigma)\mathscr
D^{IJ}_\sigma\delta(\sigma-\sigma')\\[0.2cm]
+&\frac14\left(\frac{M^{+2K}}{\rho^{\tau+2}}-\frac{M^{-2K}}{\rho^{\tau-2}}\right)
\left(\frac{\Omega^{+2K}_\sigma}{\rho^{\tau+2}}-\frac{\Omega^{-2K}_\sigma}{\rho^{\tau-2}}\right)(\sigma')\mathscr D^{IJ}_\sigma\delta(\sigma-\sigma')\\[0.2cm]
+&\frac14\left(\mathscr D^{IK}_\sigma\delta^{JL}-\mathscr
D^{JK}_{\sigma'}\delta^{IL}\right)
\left(\frac{M^{+2K}}{\rho^{\tau+2}}-\frac{M^{-2K}}{\rho^{\tau-2}}\right)\left(\frac{\Omega^{+2L}_\sigma}{\rho^{\tau+2}}-\frac{\Omega^{-2L}_\sigma}{\rho^{\tau-2}}\right)(\sigma)\delta(\sigma-\sigma')\\[0.2cm]
-&\mathscr D^{IK}_\sigma\mathscr
D^{JL}_{\sigma'}\frac{M^{KL}}{\rho^{\tau+2}\rho^{\tau-2}}(\sigma)\delta(\sigma-\sigma'),
\end{array}
\end{equation}
\begin{equation}
\begin{array}{rl}
\{M^{+2I}(\sigma),\Delta^J_k(\sigma')\}=&\frac{(k-1)\rho^{\tau+2}}{\alpha'}\delta^{IJ}\delta(\sigma-\sigma')-2\delta^{IJ}\Delta^{+2}_k\delta(\sigma-\sigma')\\[0.2cm]
-&\left[\frac{\widetilde
M^{+2-2}}{2\rho^{\tau-2}}-\frac{(\rho^{\tau+2}P^{-2}_\tau+\rho^{\tau-2}P^{+2}_\tau)}{\rho^{\tau-2}}\right](\sigma)
\mathscr D^{IJ}_\sigma\delta(\sigma-\sigma')\\[0.2cm]
+&\mathscr D^{JK}_{\sigma'}\frac{M^{IK}}{\rho^{\tau-2}}(\sigma)\delta(\sigma-\sigma')\\[0.2cm]
-&\frac12(\delta^{IJ}\delta^{KL}-\delta^{IK}\delta^{JL})\left(\frac{M^{+2K}}{\rho^{\tau+2}}-\frac{M^{-2K}}{\rho^{\tau-2}}\right)
\Omega^{+2L}_\sigma\delta(\sigma-\sigma'),
\end{array}
\end{equation}
\begin{equation}
\begin{array}{rl}
\{M^{-2I}(\sigma),\Delta^J_k(\sigma')\}=&\frac{(k+1)\rho^{\tau-2}}{\alpha'}\delta^{IJ}\delta(\sigma-\sigma')-2\delta^{IJ}\Delta^{-2}_k\delta(\sigma-\sigma')\\[0.2cm]
-&\left[\frac{\widetilde M^{+2-2}}{2\rho^{\tau+2}}+\frac{(\rho^{\tau+2}P^{-2}_\tau+\rho^{\tau-2}P^{+2}_\tau)}{\rho^{\tau+2}}\right](\sigma)\mathscr D^{IJ}_\sigma\delta(\sigma-\sigma')\\[0.2cm]
+&\mathscr D^{JK}_{\sigma'}\frac{M^{IK}}{\rho^{\tau+2}}(\sigma)\delta(\sigma-\sigma')\\[0.2cm]
+&\frac12(\delta^{IJ}\delta^{KL}-\delta^{IK}\delta^{JL})\left(\frac{M^{+2K}}{\rho^{\tau+2}}-\frac{M^{-2K}}{\rho^{\tau-2}}\right)\Omega^{-2L}_\sigma\delta(\sigma-\sigma').
\end{array}
\end{equation}
P.B. of the constraints (\ref{deltai}) and constraints (\ref{delta-2k}), (\ref{delta+2k}), (\ref{rhotau}) equal
\begin{equation}
\begin{array}{rl}
\{\Delta^I_k(\sigma),\Delta^{-2}_{(+)}(\sigma')\}=&\Omega^{-2I}_\sigma\frac{\Delta^{+2}_{k}}{\rho^{\tau+2}}\delta(\sigma-\sigma')+\frac12\mathscr D^{IJ}_\sigma(\frac12\delta^{JK}\widetilde M^{+2-2}+M^{JK})\frac{\Omega^{-2K}_\sigma}{\rho^{\tau+2}\rho^{\tau-2}}\delta(\sigma-\sigma')\\[0.2cm]
+&\frac{1}{4\rho^{\tau+2}}(\Omega^{+2I}_\sigma\Omega^{-2J}_\sigma+\Omega^{-2I}_\sigma\Omega^{+2J}_\sigma)
\left(\frac{M^{+2J}}{\rho^{\tau+2}}-\frac{M^{-2J}}{\rho^{\tau-2}}\right)\delta(\sigma-\sigma')\\[0.2cm]
+&\frac14\mathscr D^{IJ}_\sigma\frac{1}{\rho^{\tau+2}}
\left[2\Delta^J_{(+)}-\left(\mathscr
D^{JK}_\sigma\left(\frac{M^{+2K}}{\rho^{\tau+2}}-\frac{M^{-2K}}{\rho^{\tau-2}}\right)\right)\right](\sigma)\delta(\sigma-\sigma'),
\end{array}
\end{equation}
\begin{equation}
\begin{array}{rl}
\{\Delta^I_k(\sigma),\Delta^{+2}_{(-)}(\sigma')\}=&\Omega^{+2I}_\sigma\frac{\Delta^{-2}_{k}}{\rho^{\tau-2}}\delta(\sigma-\sigma')+\frac12\mathscr D^{IJ}_\sigma(-\frac12\delta^{JK}\widetilde M^{+2-2}+M^{JK})\frac{\Omega^{+2K}_\sigma}{\rho^{\tau+2}\rho^{\tau-2}}\delta(\sigma-\sigma')\\[0.2cm]
-&\frac{1}{4\rho^{\tau-2}}(\Omega^{+2I}_\sigma\Omega^{-2J}_\sigma+\Omega^{-2I}_\sigma\Omega^{+2J}_\sigma)
\left(\frac{M^{+2J}}{\rho^{\tau+2}}-\frac{M^{-2J}}{\rho^{\tau-2}}\right)\delta(\sigma-\sigma')\\[0.2cm]
+&\frac14\mathscr D^{IJ}_\sigma\frac{1}{\rho^{\tau-2}}
\left[2\Delta^J_{(-)}+\left(\mathscr
D^{JK}_\sigma\left(\frac{M^{+2K}}{\rho^{\tau+2}}-\frac{M^{-2K}}{\rho^{\tau-2}}\right)\right)
\right](\sigma)\delta(\sigma-\sigma'),
\end{array}
\end{equation}
\begin{equation}
\{\Delta^I_k(\sigma),P^{\pm2}_\tau(\sigma')\}=-{\textstyle\frac12}\mathscr D^{IJ}_\sigma{\textstyle\frac{M^{\mp2J}}{(\rho^{\tau\mp2})^2}}\delta(\sigma-\sigma').
\end{equation}
$so(1,9)/(so(1,1)\times so(8))$ coset generators satisfy the
following P.B. with the scalar second-class constraints
\begin{equation}
\{M^{+2I}(\sigma),\Delta^{+2}_{k}(\sigma')\}={\textstyle\frac{1}{\rho^{\tau-2}}}\left[\delta^{IJ}({\textstyle\frac12}\widetilde
M^{+2-2}-\rho^{\tau+2}P^{-2}_\tau)-M^{IJ}\right]\Omega^{+2J}_\sigma\delta(\sigma-\sigma'),
\end{equation}
\begin{equation}
\{M^{-2I}(\sigma),\Delta^{-2}_{k}(\sigma')\}=-{\textstyle\frac{1}{\rho^{\tau+2}}}\left[\delta^{IJ}({\textstyle\frac12}\widetilde
M^{+2-2}+\rho^{\tau-2}P^{+2}_\sigma)+M^{IJ}\right]\Omega^{-2J}_\sigma\delta(\sigma-\sigma'),
\end{equation}
\begin{equation}
\begin{array}{c}
\{M^{+2I}(\sigma),\Delta^{-2}_k(\sigma')\}=-\left[\Delta^I_k+\Omega^{-2I}_\sigma
P^{+2}_\tau+\frac12\left(\mathscr D^{IJ}_\sigma\left(\frac{M^{+2J}}{\rho^{\tau+2}}+\frac{M^{-2J}}{\rho^{\tau-2}}\right)\right)\right]\delta(\sigma-\sigma')\\[0.3cm]
+(\mathscr
D^{IJ}_\sigma+\frac12\Omega^{+2-2}_\sigma\delta^{IJ})\frac{M^{+2J}}{\rho^{\tau+2}}(\sigma)\delta(\sigma-\sigma'),
\end{array}
\end{equation}
\begin{equation}
\begin{array}{c}
\{M^{-2I}(\sigma),\Delta^{+2}_k(\sigma')\}=-\left[\Delta^I_k+\Omega^{+2I}_\sigma
P^{-2}_\tau+\frac12\left(\mathscr D^{IJ}_\sigma\left(\frac{M^{+2J}}{\rho^{\tau+2}}+\frac{M^{-2J}}{\rho^{\tau-2}}\right)\right)\right]\delta(\sigma-\sigma')\\[0.3cm]
+(\mathscr
D^{IJ}_\sigma-\frac12\Omega^{+2-2}_\sigma\delta^{IJ})\frac{M^{-2J}}{\rho^{\tau-2}}(\sigma)\delta(\sigma-\sigma').
\end{array}
\end{equation}
Scalar second-class constraints are characterized by the P.B. relations
\begin{equation}
\begin{array}{rl}
\{\Delta^{-2}_k(\sigma),\Delta^{-2}_{k'}(\sigma')\}=&\frac{(k-k')}{2\rho^{\tau+2}}\Omega^{-2I}_\sigma\Phi^I\delta(\sigma-\sigma'),\\[0.2cm]
\{\Delta^{+2}_k(\sigma),\Delta^{+2}_{k'}(\sigma')\}=&\frac{(k-k')}{2\rho^{\tau-2}}\Omega^{+2I}_\sigma\Phi^I\delta(\sigma-\sigma'),\\[0.2cm]
\{\Delta^{+2}_k(\sigma),\Delta^{-2}_{k'}(\sigma')\}=&
\frac{1}{2\rho^{\tau+2}\rho^{\tau-2}}\Omega^{-2I}_\sigma\Omega^{+2J}_\sigma(\frac12\delta^{IJ}\widetilde
M^{+2-2}+M^{IJ})\delta(\sigma-\sigma'),\\[0.2cm]
\{\Delta^{\pm2}_{(\mp)}(\sigma),P^{\mp2}_\tau(\sigma')\}=&\pm\frac{1}{\alpha'}\delta(\sigma-\sigma'),\\[0.2cm]
\{\Delta^{\pm2}_{(\mp)}(\sigma),P^{\pm2}_\tau(\sigma')\}=&\frac{1}{2(\rho^{\tau\mp})^2}\Omega^{\pm2I}_\sigma
M^{\mp2I}\delta(\sigma-\sigma').
\end{array}
\end{equation}
Fermionic second-class constraints (\ref{fermi2}) satisfy P.B.
relations among themselves
\begin{equation}
\begin{array}{rl}
\{\widetilde D^-_A(\sigma),\widetilde
D^-_B(\sigma')\}=&-\frac{i\rho^{\tau-2}}{4\alpha'}\delta_{AB}\delta(\sigma-\sigma')
+\frac{i}{4}\delta_{AB}\left[\Delta^{-2}_{(+)}-\partial_\sigma
P^{-2}_\tau+\frac12\Omega^{+2-2}_\sigma
P^{-2}_\tau\right.\\[0.2cm]
+&\left.\frac12\Omega^{-2I}_\sigma\left(\frac{M^{-2I}}{\rho^{\tau-2}}-\frac{M^{+2I}}{\rho^{\tau+2}}\right)\right]\delta(\sigma-\sigma'),
\end{array}
\end{equation}
and with bosonic second-class constraints
\begin{equation}
\begin{array}{rl}
\{\widetilde D^-_A(\sigma),\Delta^I_k(\sigma')\}=&
\frac{i}{4\rho^{\tau-2}}\gamma^I_{A\dot A}\mathcal
D_\sigma\eta^-_{\dot
A}\Delta^{-2}_k\delta(\sigma-\sigma')\\[0.2cm]
+&\frac{i}{16\rho^{\tau-2}}\left(\Omega^{-2I}_\sigma\gamma^J_{A\dot
A}\mathcal D_\sigma\eta^-_{\dot
A}+\Omega^{-2J}_\sigma\gamma^I_{A\dot A}\mathcal
D_\sigma\eta^-_{\dot A}\right)
\left(\frac{M^{-2J}}{\rho^{\tau-2}}-\frac{M^{+2J}}{\rho^{\tau+2}}\right)\delta(\sigma-\sigma')\\[0.2cm]
+&\mathscr
D^{IJ}_{\sigma'}\left[\frac{1}{2\rho^{\tau+2}}\gamma^{J}_{A\dot
A}\widetilde D^+_{\dot
A}-\frac{i}{8\rho^{\tau+2}\rho^{\tau-2}}\left(\frac12\delta^{JK}\widetilde
M^{+2-2}-M^{JK}\right)\gamma^K_{A\dot A}\mathcal
D_\sigma\eta^-_{\dot A}\right.\\[0.2cm]
+&\frac{i}{8\rho^{\tau+2}\rho^{\tau-2}}\mathcal D_\sigma\eta^+_A
M^{-2J} \left.-\frac{i}{8\rho^{\tau-2}}\gamma^{J}_{A\dot
A}\mathcal D_\sigma\eta^-_{\dot
A}P^{-2}_\tau\right](\sigma)\delta(\sigma-\sigma'),
\end{array}
\end{equation}
\begin{equation}
\begin{array}{rl}
\{\widetilde
D^-_A(\sigma),M^{+2I}(\sigma')\}=&-\left(\gamma^I_{A\dot
A}\widetilde D^+_{\dot A}
-\frac{i}{4\rho^{\tau-2}}(\frac12\delta^{IJ}\widetilde
M^{+2-2}-M^{IJ})\gamma^J_{A\dot A}\mathcal D_\sigma\eta^-_{\dot
A}\right)\delta(\sigma-\sigma')\\[0.2cm]
-&\frac{i}{4\rho^{\tau-2}}\left(\mathcal D_\sigma\eta^+_A
M^{-2I}+\rho^{\tau+2}\gamma^{I}_{A\dot
A}\mathcal D_\sigma\eta^-_{\dot
A}P^{-2}_\tau\right)\delta(\sigma-\sigma'),
\end{array}
\end{equation}
\begin{equation}
\begin{array}{rl}
\{\widetilde
D^-_A(\sigma),\Delta^{+2}_k(\sigma')\}=&\frac{i(k+1)}{8\alpha'}\mathcal
D_\sigma\eta^+_A\delta(\sigma-\sigma')
+\frac{i}{8\rho^{\tau-2}}\gamma^{I}_{A\dot A}\mathcal
D_\sigma\eta^-_{\dot A}\left[\Delta^I_k+\Omega^{+2I}_\sigma
P^{-2}_\tau\right.\\[0.2cm]
+&\left.\frac12\mathscr
D^{IJ}_\sigma\left(\frac{M^{+2J}}{\rho^{\tau+2}}-\frac{M^{-2J}}{\rho^{\tau-2}}\right)\right]\delta(\sigma-\sigma'),
\end{array}
\end{equation}
\begin{equation}
\{\widetilde D^-_A(\sigma),P^{+2}_\tau(\sigma')\}=-{\textstyle\frac{i}{8(\rho^{\tau-2})^2}}\gamma^I_{A\dot
A}\mathcal D_\sigma\eta^-_{\dot A}M^{-2I}\delta(\sigma-\sigma').
\end{equation}

\end{document}